\documentclass[12pt]{article}
\usepackage{amssymb,amsmath}
\usepackage{color,graphicx}
\usepackage{subcaption}
\usepackage{cite}
\newcommand\fverb{\setbox\pippobox=\hbox\bgroup\verb}
\newcommand\fverbdo{\egroup\medskip\noindent%
                              \fbox{\unhbox\pippobox}\ }
\newcommand\fverbit{\egroup\item[\fbox{\unhbox\pippobox}]}
\newbox\pippobox

\newcommand{\be} {\begin{equation}}
\newcommand{\ee} {\end{equation}}
\newcommand{\beq} {\begin{equation}}
\newcommand{\eeq} {\end{equation}}
\newcommand{\bea}{\begin{eqnarray}}
\newcommand{\eea}{\end{eqnarray}}
\newcommand{\bear}{\begin{eqnarray}}
\newcommand{\eear}{\end{eqnarray}}

\newcommand{\rc}{\nonumber\\}
\newcommand{\Tr}{\mbox{Tr}}    

\begin{document}

\begin{flushright}
HIP-2018-16/TH
\end{flushright}

\begin{center}

\centerline{\Large {\bf Low-energy modes in anisotropic holographic fluids}}

\vspace{8mm}

\renewcommand\thefootnote{\mbox{$\fnsymbol{footnote}$}}
Georgios Itsios,${}^{1}$\footnote{gitsios@gmail.com}
Niko Jokela,${}^{2,3}$\footnote{niko.jokela@helsinki.fi}
Jarkko J\"arvel\"a,${}^{2,3}$\footnote{jarkko.jarvela@helsinki.fi}
and Alfonso V. Ramallo${}^{4,5}$\footnote{alfonso@fpaxp1.usc.es}

\vspace{4mm}
${}^1${\small \sl Instituto de F\'is\'ica Te\'orica, UNESP-Universidade Estadual Paulista,} \\
{\small \sl R. Dr. Bento T. Ferraz 271, Bl. II,} \\
{\small \sl Sao Paulo 01140-070, SP, Brazil} 

\vskip 0.2cm
${}^2${\small \sl Department of Physics} and ${}^3${\small \sl Helsinki Institute of Physics} \\
{\small \sl P.O.Box 64} \\
{\small \sl FIN-00014 University of Helsinki, Finland} 

\vskip 0.2cm
${}^4${\small \sl Departamento de  F\'\i sica de Part\'\i  culas} \\
{\small \sl Universidade de Santiago de Compostela} \\
{\small \sl and} \\
${}^5${\small \sl Instituto Galego de F\'\i sica de Altas Enerx\'\i as (IGFAE)} \\
{\small \sl E-15782 Santiago de Compostela, Spain} 

\end{center}

\vspace{8mm}
\numberwithin{equation}{section}
\setcounter{footnote}{0}
\renewcommand\thefootnote{\mbox{\arabic{footnote}}}

\begin{abstract}
\noindent
In this work we will study the low-energy collective behavior of spatially anisotropic dense fluids in four spacetime dimensions. We will embed a massless flavor D7-brane probe in a generic geometry which has a metric possessing anisotropy in the spatial components. We work out generic formulas of the low-energy excitation spectra and two-point functions for charged excitations at finite baryon chemical potential. In addition, we specialize to a certain Lifshitz geometry and discuss in great detail the scaling behavior of several different quantities.

\end{abstract}

\newpage
\tableofcontents
\newpage

\section{Introduction}
Holography has shown to be a useful tool to study various gauge field theories particularly at finite density \cite{CasalderreySolana:2011us,Brambilla:2014jmp,Zaanen:2015oix,Hartnoll:2016apf}. The field has matured to the point where the low hanging fruit has been picked, while the more complicated problems have been sitting aside until recent years. One physically very relevant situation, where a complicated problem occurs, is a system where some or all of the spacetime symmetries are broken. Even in one of the simplest cases, where the translational symmetry is spontaneously broken in only one of the field theory directions leads to tedious calculations and numerical work.
However, despite involved numerics, there is already a vast literature addressing holographic striped phases \cite{Ooguri:2010kt, Ooguri:2010xs, Bayona:2011ab, Bergman:2011rf, Donos:2012gg, Jokela:2012vn, BallonBayona:2012wx, Bu:2012mq, Jokela:2012se, Rozali:2012es, Donos:2012yu,Donos:2013wia, Withers:2013loa, Withers:2013kva, Rozali:2013ama, Ling:2014saa, Jokela:2014dba, Donos:2016hsd, Amoretti:2016bxs,Cremonini:2017usb,Jokela:2017fwa} and in particular the conductivities of charged excitations \cite{Jokela:2016xuy,Cremonini:2017qwq,Jokela:2017ltu,Amoretti:2017frz,Amoretti:2017axe,Donos:2018kkm,Gouteraux:2018wfe,Li:2018vrz}.

Another way of breaking the symmetries is to maintain homogeneity but singling out some of the spatial directions and study anisotropic situations. The dual gravity backgrounds for many field theories possessing anisotropy have been constructed \cite{Azeyanagi:2009pr,Mateos:2011ix,Mateos:2011tv,Giataganas:2012zy,Rougemont:2014efa,Fuini:2015hba,Conde:2016hbg,Gursoy:2016ofp,Penin:2017lqt,Bea:2017iqt}. Given the fact that anisotropic backgrounds are in many ways computationally much tamer than inhomogeneous ones, surprisingly, not much has been said about the transport of excitation spectra or conductivities, for a recent example see, however, \cite{Giataganas:2017koz}. 

In this paper, we set out to study a family of anisotropic backgrounds with generic metric components. We introduce a probe D7-brane thus adding fundamental degrees of freedom in the otherwise dual pure-glue field theories. We are particularly interested in low-energy excitations and transport properties of fundamental degrees of freedom. The backgrounds we have in mind have only been constructed numerically in the literature, and our aim in this paper is not to reconstruct them. Instead, upon generalizing the methods developed in \cite{Jokela:2015aha,Itsios:2015kja,Itsios:2016ffv}, we leave our results in forms that are directly applicable if numerical backgrounds are plugged in. Here we are content with discussing general behavior with varying amounts of density. However, we do cross-check our analytic formulas numerically in certain Lifshitz geometries and find excellent agreement.
  
We continue with the introduction by a technical review of the dual gravity setups that are relevant in the present case and we will put them in a broader physical context. Here we are going to be a bit cavalier about relating the conventions in different contexts and write down the formulas as in the original papers. However, in the rest of the paper we pay careful attention to all the numerical factors and keep consistent conventions and will in particular relate the two metrics discussed below.

Most of the backgrounds with anisotropy found so far are inspired by the one obtained in \cite{Azeyanagi:2009pr}, which corresponds to a system of D3- and D7-branes. The latter are completely dissolved in the geometry and induce a RR axion $\chi$ which is linear in one of the spatial Minkowski coordinates.  In section 3 of \cite{Azeyanagi:2009pr} the authors found a running solution which interpolates between the scaling solution in the deep IR and the $AdS_5 \times S^5$ geometry in the UV. This solution has zero temperature and was found numerically. They also found, in section 7, a scaling D4-D6 solution. In this case there are two anisotropic directions $w^1$ and $w^2$ and the anisotropy is induced by a RR two-form $F_2 \propto dw^1 \wedge dw^2$ . In appendix B they work out a ten-dimensional Lifshitz solution corresponding to a D3-D5 system with F-string sources. This solution is written in eq. (8.1) of \cite{Azeyanagi:2009pr} and corresponds to a dynamical exponent $z = 7$.

A more general class of anisotropic gravity solutions was found in \cite{Mateos:2011ix,Mateos:2011tv}.  This solution has non-zero temperature in general and generalizes the running solution of \cite{Azeyanagi:2009pr}. The solution found in \cite{Mateos:2011ix,Mateos:2011tv} is numeric, although there are some analytic expressions for the functions in some limits. For example, there are expressions for small anisotropic parameter (actually for $a \ll T$) (see appendix D in \cite{Mateos:2011tv}). There are also expressions near the boundary (section 3) and for small $T \ll a$ (appendix E).

The anisotropic background of Mateos-Trancanelli (MT) has been used, for example, in \cite{Patino:2012py,Jahnke:2013rca} to study the thermal photon production in a plasma. They embedded flavor D7-brane probes in the MT background and analyzed the fluctuations of the worldvolume gauge field (at zero charge
density). The goal was to obtain the current-current correlators for photons with $k_0 = |\vec k|$ and to get the photon production rate for different angles and energies. In \cite{Patino:2012py} the quarks are massless, while in \cite{Jahnke:2013rca} the embedding of the flavor branes corresponds to massive quarks. In \cite{Wu:2013qja} the authors considered the effect of a constant magnetic field on the photon spectrum.

In \cite{Mamo:2012sy} the author studied the running of the shear viscosity $\eta$ in the MT solution. The anisotropy induces a dependence of the shear viscosity with the scale, which manifests itself
in a temperature dependence of $\eta$ which violates the lower bound on the shear viscosity to the entropy ratio bound \cite{Kovtun:2004de}. 
In \cite{Bu:2014cca} the author analyzed the Chern-Simons diffusion rates for the MT solution. To obtain this quantity one has to analyze the fluctuations of the axion field $\chi$.
  
The papers \cite{Cheng:2014qia} and \cite{Cheng:2014sxa} deal with a generalization of the MT solution to the case in which a $U(1)$ gauge field is added. This corresponds to an R-charge chemical potential. The black holes constructed are of the Reissner-Nordstr\"om type. In this solution the internal $S^5$ is deformed, which corresponds to new internal components of the RR five-form $F_5$. These two papers mimic the MT one, but the solution depends on an additional charge parameter $Q$, in addition to the axionic parameter $a$. They also address the thermodynamics in the presence of the chemical potential. The conductivities of this background were derived in \cite{Ge:2014aza}.

The MT approach is of course not the only one to generate anisotropy in holography. Indeed, already some time before MT, the authors of \cite{Janik:2008tc} found a solution of Einstein's equation corresponding to an anisotropic energy momentum tensor (with two pressures) and they analyzed the quasinormal modes for R-charge diffusion. The recent paper \cite{Banks:2015aca} obtained a new solution which, apart from the dilaton and axion, has an extra scalar field $X$. They argued that this new solution is thermodynamically preferred over the MT one at low temperatures.

In an interesting and detailed paper \cite{Jain:2014vka} a new anisotropic solution was found. In this case the anisotropy is induced by a dilaton profile of the type $\phi = \rho z$ ($\rho$ being a constant similar to $a$ in MT). Now the metric at the IR is of the type $AdS_4 \times R$ (and not Lifshitz). The section 3 includes the analysis of the anistropic thermodynamics in this setting.

Another way of getting supergravity solutions with anisotropy is by considering backgrounds dual to non-commutative gauge theories. As an example of these, in \cite{Roychowdhury:2015lma}, the charge diffusion in the D1-D3 solution of Maldacena-Russo is studied. The author solves Maxwell's equation in this background and derives the longitudinal and Hall conductivities.

After this rather lengthy review of the existing literature, let us summarize our aim. We will address the problem of studying the charge transport properties of an anisotropic plasma using top-down holographic methods. Our main motivation to follow top-down approach, instead of a bottom-up approach, is that the field theory dual is well-established and the anisotropy has a well-defined origin in the gauge theory.  We will concentrate on a particular setup in which the anisotropy is introduced by a space-dependent axion, which corresponds to  ${\cal{N}}=4$,  $(3+1)$-dimensional super Yang-Mills deformed by a theta angle linearly dependent on one of the coordinates. The corresponding  supergravity backgrounds have been obtained in references \cite{Azeyanagi:2009pr} and \cite{Mateos:2011ix,Mateos:2011tv}. 

We will start discussing the background geometry which has a general metric possessing anisotropy as in \cite{Mateos:2011ix,Mateos:2011tv}. We then embed a probe D7-brane in this geometry and discuss the associated thermodynamics in Sec.~\ref{setup} to the extent possible without specifying a particular solution. We then switch to discussing the fluctuation spectra of the flavor degrees of freedom in Sec.~\ref{sec:fluctuations}. We also work out the two-point functions and make a non-trivial check of the formulas. In Sec.~\ref{sec:lifshitz} we specialize to the gravity solution found in \cite{Azeyanagi:2009pr} and evaluate the formulas laid out in the preceding section. In Sec.~\ref{sec:num} we further perform a numerical analysis and show that our analytical formulas agree with the numerics very accurately. Sec.~\ref{sec:num} contains a brief summary and an outlook of possible outgrowths of our work. Some computational details are relegated in App.~\ref{appendix:calculations}.


\section{Gravity background and flavor thermodynamics}\label{setup}

We consider the low-energy physics on a probe brane in a spatially anisotropic background. The background setup was originally studied in \cite{Mateos:2011ix,Mateos:2011tv}. The action is that of a type IIB supergravity where we only have a dilaton, an axion, and a RR five-form,
\beq
S = \frac{1}{2\kappa_{10}^2}\int {\rm d}^{10}x\sqrt{-g}\left[{\rm e}^{-2\phi}(R+4\partial_M\phi\partial^M\phi)-\frac{1}{2}F_1^2-\frac{1}{4\cdot 5!}F_5^2 \right]\ , \label{eq:action}
\eeq
where $M=0,\ldots,9$ and $F_1=d\chi$ is the axion field strength. The metric Ansatz in the string frame is 
\beq
ds^2=\frac{L_s^2}{u^2}\left(-\mathcal{F}\mathcal{B}dt^2+dx^2+dy^2+\mathcal{H}dz^2+\frac{du^2}{\mathcal{F}}\right)+L_s^2\mathcal{Z}d\Omega_{S^5}^2\ , \label{eq:metric}
\eeq
where all the components depend only on the radial coordinate $u$, which is $0$ at the UV boundary. The RR five-form is set to be self-dual and chosen to be $F_5 = \alpha(\Omega_{S^5}+\star\Omega_{S^5})$, where $\Omega_{S^5}$ is the volume element of a five-sphere  and $\alpha$ is a constant determined by flux quantization. The axion is linear in this Ansatz: $\chi = a z$. 

The authors of \cite{Mateos:2011ix,Mateos:2011tv} studied mostly solutions that are asymptotically AdS. This enables us to set some regularity conditions. Due to freedom in parametrization, we can set $\mathcal{H}(0)=\mathcal{B}(0)=1$ at the boundary along with $\phi(0)=0$. The function $\mathcal{F}$ is the blackening factor and vanishes at the horizon, $u_H$.  The equations of motion for these Ans\"atze require $\alpha = 4{\rm e}^{-\phi(0)}L_s^2=4L_s^2$ and $\mathcal{F}(0)=1/\mathcal{Z}(0)$. Finally, the Ansatz was further simplified by setting
\beq
\mathcal{H}={\rm e}^{-\phi},\quad \mathcal{Z}={\rm e}^{\frac{\phi}{2}}\ ,\label{eq:simple}
\eeq
which restricts $\mathcal{F}(0)=\mathcal{Z}(0)=1$. With these, we only need to find solutions for $\mathcal{F}$, $\mathcal{B}$, and $\phi$.

The solutions emerging with these Ans\"atze exhibit scaling behavior with pure AdS$_5$ metric at the UV boundary and it becomes more and more anisotropic closer to the horizon. The anisotropicity is controlled by the axion strength, $a^2$. These solutions can be found analytically near the UV boundary at the low- and high-temperature limits. For intermediate temperatures, only numerical results are available. Due to the difficulty in obtaining analytical solutions, most of the following calculations will be presented with the metric in \eqref{eq:metric} although in a more condensed notation. Later on, we will consider a special case of fixed-point Lifshitz metric in Sec.~\ref{sec:lifshitz}.

We now wish to embed a probe $D7$-brane into the geometry. We first find a classical solution and study its thermodynamics and then move on to the fluctuations in Sec.~\ref{sec:fluctuations}. For the metric of the five-sphere, we will be using the fibration
\beq
d\Omega_{S^5}^2 = d\theta^2+\sin^2\theta\, d\psi^2+\cos^2\theta\, d\Omega_{S^3}^2 \ .
\eeq
The probe D7-brane will span the directions $(t,x,y,z,u,\Omega_3)$. We furthermore turn on a gauge field on the brane, $F=A_t'(u) du\wedge dt$. We have chosen the gauge $A_u=0$. The brane will obey the dynamics following from the DBI action:
\beq
S_{DBI} = -T_7\int d^{8}x{\rm e}^{-\phi}\sqrt{-{\det}(g+F)}\ ,
\eeq
where the metric $g$ is the induced metric on the brane, {\emph i.e.},
\beq
ds^2_8=\frac{L_s^2}{u^2}\left(-\mathcal{F}\mathcal{B}dt^2+dx^2+dy^2+\mathcal{H}dz^2+\frac{du^2}{\mathcal{F}}\right)+L_s^2\mathcal{Z}\,\cos^2\theta\,d\Omega_{S^3}^2 \ .
\eeq
Moreover, notice that we choose to absorb the factors of $2\pi\alpha'$ in the definitions of the gauge fields. We only focus on massless fundamentals, so it is consistent to integrate over the internal directions as the embedding does not vary inside $\Omega_3$. In other words, $\psi={\mathrm{const.}}$ and $\theta=0$ are consistent solutions to the equations of motion.

We choose the following conventions for the use of indices. Lower case Latin letters $i,j,\ldots$ denote spatial directions. Greek letters $\mu,\nu,\ldots$ correspond to Poincar\'e coordinates or the coordinates along the boundary. Capital Latin letters $A,B,\ldots$ correspond to all the coordinates of the metric. In addition, prime indicates differentiation with respect to $u$.

The action evaluates to 
\beq
S_{DBI} = -T_7V\Omega_3L_s^3\int {\rm d}u\,{\rm e}^{-\phi}g_{xx}\sqrt{g_{zz}\mathcal{Z}^3(|g_{tt}|g_{uu}-{A'_t}^2)}\ .
\eeq
The  $\Omega_3$ is the volume of the three-sphere while $V$ is the 4-volume of the space spanned by $t$, $x$, $y$, and $z$.

We see that $A_t$ is a cyclic variable so it can be easily solved from the Euler-Lagrange equations to give
\beq
A'_t = d\frac{\sqrt{|g_{tt}|g_{uu}}}{\sqrt{W+d^2} } \ , 
\qquad\qquad
\ W \equiv {\rm e}^{-2\phi}L_s^6\mathcal{Z}^3g_{xx}^2g_{zz} \ .
\eeq
Here, $d$ is a constant and we will show that it is proportional to the particle density.

The chemical potential is
\beq
\mu = \int\limits_{0}^{u_H} {\rm d}u\,A'_t \ ,
\eeq
while the particle density is 
\beq
\rho = -\frac{1}{V}\left(\frac{\partial\Omega_{grand}}{\partial \mu}\right)_{V,T} = -\frac{1}{V}\left(\frac{\partial d}{\partial \mu}\right)_{V,T}\left(\frac{\partial\Omega_{grand}}{\partial d}\right)_{V,T}=T_7 \Omega_3 d \ ,
\eeq
where $\Omega_{grand}=-S_{\rm on-shell}$ is the unregularized grand potential,
\beq
\Omega_{grand} = T_7 V\Omega_3L_s^{14} \int\limits_{0}^{u_H}{\rm d}u\,\frac{\sqrt{|g_{tt}|g_{uu}}W}{\sqrt{d^2+W}}=T_7V\Omega_3\int\limits_{0}^{u_H}{\rm d}u\,{\rm e}^{-\frac{3\phi}{4}}u^{-5}\sqrt{\frac{\mathcal{B}}{L_s^{12}+d^2u^6{\rm e}^{\frac{3}{2}\phi}}} \ .
\eeq

For further thermodynamic computations, the on-shell action needs to be regularized. The simplest way for this is to subtract the $0$-density action,
\beq
S^{(\rm on-shell)}_{{\rm reg.}}=-T_7 V\Omega_3 L_s^3\int\limits_{0}^{u_H}{\rm d}u\,\sqrt{|g_{tt}|g_{uu}}\left(\frac{W}{\sqrt{d^2+W}}-\sqrt{W}\right).
\eeq
If $d$ was small, we would also need to consider the contributions from the $d=0$ solution. The proper regularization of the $d=0$ action has been done in \cite{Avila:2016mno}. We only care about energy differences, so this suffices to our needs.

The temperature is given by the formula
\beq
T_H = \frac{\sqrt{\mathcal{B}(u_H)}|\mathcal{F}'(u_H)|}{4\pi}.
\eeq

The energy density can be found with a Legendre transformation $\epsilon = \frac{\Omega_{grand}}{V}+\mu\rho$. In order to determine the pressures along  the different directions, let us put the system in a box of sides $L_x$, $L_y$ and $L_z$. Then, the pressure along direction $q$ can be found with 
\beq
p_q = -\frac{L_q}{V}\frac{\partial \Omega_{grand} }{\partial L_q}\,\,,
\qquad \qquad q=x,y,z\,\,.
\eeq
The computation of pressures seems like a trivial task, after all, all the unknown quantities depend only on $u$. However, the $z$ direction is a special case and the action depends non-trivially on $L_z$ as seen in the explicit Lifshitz scaling solution of section \ref{sec:lifshitz}. With these, we can also compute the speed of sound with
\beq
c_q^2 = \left(\frac{\partial p_q}{\partial \epsilon}\right)_s \ .
\eeq
Later in Sec.~\ref{sec:lifshitz} we will explicitly evaluate all the formulas in this section in a particular geometry.

\section{Low-energy modes}\label{sec:fluctuations}
We now move on to considering the spectrum of fluctuations of the gauge fields. We modify the action $g+F\to (g+F+f)$ and expand to second order in $f$ where $f_{\mu\nu}=\partial_{\mu}a_{\nu}-\partial_{\nu}a_{\mu}$. To expand the action, we need to expand the determinant
\beq
\sqrt{-{\rm det}(g+F+f)}=\sqrt{-{\rm det}(g+F)}\sqrt{{\rm det}(1+(g+F)^{-1}f)}\ .
\eeq
We can use the following expansion
\beq
\sqrt{{\rm det}(1+X)}=1+\frac{\Tr X}{2}+\frac{(\Tr X)^2}{8}-\frac{\Tr X^2}{4}+\mathcal{O}(X^3)\ .
\eeq  
The first order terms vanish as we are fluctuating around a saddle point. The inverse of $(g+F)$ is
\bear
(g+F)^{-1}&=&\left[\begin{array}{ccccc}\frac{-g_{uu}}{g_{uu}|g_{tt}|-{A'_t}^2} & \frac{A'_t}{|g_{tt}|g_{uu}-{A'_t}^2} &0& 0 & \cdots \\ 
\frac{-A'_t}{|g_{tt}|g_{uu}-{A'_t}^2} & \frac{|g_{tt}|}{g_{uu}|g_{tt}|-{A'_t}^2} & 0 &0& \cdots \\
0 & 0 & g_{zz}^{-1} &0&\cdots \\
0&0&0&g_{xx}^{-1} &\cdots\\
\vdots &\vdots&\vdots&\vdots& \ddots\end{array}\right]\rc
 &=& G + J\ ,
\eear
where $G$ is the diagonal part while $J$ is the antisymmetric part of the matrix. In the above matrix, $t$ is the first coordinate, $u$ the second, $z$ the third etc.

The second order term of the determinant is
\beq\label{eq:lagrangian2order}
\sqrt{-{\rm det} (g+F+f)}_{\rm{2nd}\,\rm{ord.}}=\sqrt{|g_{tt}|g_{uu}-{A'_t}^2}\left(\frac{J^{AB}J^{CD} }{8} - \frac{G^{AD}G^{BC}+J^{DA}J^{BC}}{4} \right)f_{AB}f_{CD}\ .
\eeq
We get the equations of motion from the Euler-Lagrange equations. We make the assumption that $a_{\mu}$ do not depend on the spherical coordinates. The equations are
\beq
\partial_N \frac{W\sqrt{|g_{tt}|g_{uu}}}{\sqrt{W+d^2}}\left(\frac{J^{NM}J^{AB}}{2}-G^{MA}G^{BN}-J^{BN}J^{MA} \right) f_{AB}=0 
\eeq
for all $M$. It turns out that in our case when only the $tu$ components of $J$ are non-zero, the $J$ matrices do not contribute to either the action nor the equations of motion. They will not appear in the rest of the calculations.

Our gauge condition is $a_u=0$. We get an important constraint equation from this by setting $m=u$
\bear
&&G^{tt}\partial_t a'_t+\sum\limits_{i}^{}G^{ii}\partial_i a'_i=0\,\,\rc
&\Leftrightarrow& \partial_t a'_t -\sum_i v_i^2\partial_i a'_i=0, \quad v_i^2 = -\frac{G^{ii}}{G^{tt}}\ .
\eear
For spatial or temporal coordinates, we have
\beq
a''_{\mu}+ \partial_u\log\left(\sqrt{|g_{tt}|g_{uu}}\frac{W}{\sqrt{W+d^2}}G^{\mu\mu}G^{uu}\right)a'_{\mu}-\sum\limits_{\lambda}^{}\frac{G^{\lambda\lambda}}{G^{uu}}\partial_{\lambda}f_{\mu\lambda}=0\ .
\eeq

We make a further assumption. Due to rotational invariance in the $xy$-plane, we assume that $a_{\mu}$'s are independent of $y$. Then we Fourier transform along directions $t$, $x$, and $z$ with
\beq
a_{\mu}(u,t,x,z)=\int_{{\mathbb R}^3}{\rm d}\omega {\rm d}k_{x}{\rm d}k_z\,a_{\mu}(u,\omega,{\vec k}){\rm e}^{i({\vec k}\cdot (x,z)-\omega t)}\ .
\eeq
The equations of motion as functions of $\omega$ and ${\vec k}=(k_x,k_z)$ are
\bear
\label{eq:as}
&&\!\!\!\!\!\!\!\omega a'_t + (k_x v_x^2a'_x+k_z v_z^2 a'_z)=0\,\rc
&&\!\!\!\!\!\!\!a''_t+\partial_u\log\left(\sqrt{g_{uu}|g_{tt}|}\frac{W}{\sqrt{W+d^2}} G^{uu}G^{tt}\right)a'_t +\frac{G^{tt}}{G^{uu}}\left(v_x^2k_xE_x +v_z^2k_zE_z \right)=0\,\rc
&&\!\!\!\!\!\!\!a''_x+\partial_u\log\left(\sqrt{g_{uu}|g_{tt}|}\frac{W}{\sqrt{W+d^2}}G^{uu}G^{xx} \right)a'_x 
\,\rc
&&\quad\quad\quad\quad\quad\quad\quad-\frac{G^{tt}}{G^{uu}}\left(\omega E_x+\frac{k_z v_z^2}{\omega}(k_x\, E_z-k_z\, E_x) \right)=0\,,\rc
&&\!\!\!\!\!\!\!a''_z+\partial_u\log\left(\sqrt{g_{uu}|g_{tt}|}\frac{W}{\sqrt{W+d^2}} G^{uu}G^{zz}\right)a'_z \nonumber\\&&\quad\quad\quad\quad\quad\quad\quad-\frac{G^{tt}}{G^{uu}}\left(\omega E_z-\frac{k_xv_x^2}{\omega}(k_x\, E_z-k_z\, E_x)  \right)=0\,,
\eear
where we have already used gauge-invariant quantities
\beq
E_x = \omega a_x+k_x\, a_t,\quad E_z = \omega a_z+k_z\, a_t \ .
\eeq
Using the gauge constraint and definitions of $E_x$ and $E_z$, we can solve the derivatives of $a_{\mu}$ in terms of $E_x$ and $E_z$
\bear
a'_t &=& \frac{1}{-\omega^2+(v_x^2k_x^2 +v_z^2k_z^2)}\left(v_x^2k_x\,E'_x+v_z^2k_z\,E'_z\right)
\,\rc\rc
a'_x &=& \frac{-1}{-\omega^2+(v_x^2k_x^2 +v_z^2k_z^2)}\left(\left(\omega -\frac{k_z^2}{\omega}v_z^2\right)E'_x +\frac{k_xk_z}{\omega}v_z^2 E'_z\right)\,\rc\rc
a'_z &=& \frac{-1}{-\omega^2+(v_x^2k_x^2 +v_z^2k_z^2)}\left(\left(\omega-\frac{k_x^2}{\omega} v_x^2\right)E'_z+\frac{k_xk_z}{\omega}v_x^2 E'_x\right)\,\,.
\eear

Plugging these into the above equations of motion, we can express the equations of motion in terms of the gauge-invariant fields. Using suitable linear combinations, we can find equations that have the 2nd derivative on only one field. For $E_x$ it is
\bear
&&\!\!\!\!\!E''_x+\left(\partial_u{\log\left(\sqrt{g_{uu}|g_{tt}|}\frac{W}{\sqrt{W+d^2}}G^{uu}G^{xx} \right)}+\frac{\partial_u\log(v_x^2)k_x^2v_x^2}{\omega^2-(v_x^2k_x^2+v_z^2k_z^2)}\right)E'_x\rc\rc
&&\quad\quad\quad\quad\quad\quad+\frac{k_xk_z v_z^2\partial_u\log(v_x^2)E'_z}{\omega^2-(v_x^2k_x^2+v_z^2k_z^2)} - \frac{G^{tt}}{G^{uu}}\left(\omega^2-(v_x^2k_x^2+v_z^2k_z^2)\right)E_x=0 
\,\,,
\eear
and for $E_z$ it is
\bear
&&\!\!\!\!\!E_z'' + \left(\partial_u\log\left(\sqrt{g_{uu}|g_{tt}|}\frac{W}{\sqrt{W+d^2}}G^{uu}G^{zz} \right) +\frac{\partial_u\log(v_z^2)k_z^2 v_z^2}{\omega^2-(k_z^2 v_z^2+k_x^2v_x^2)}\right)E_z'\rc\rc
&&\quad\quad\quad\quad+\frac{k_xk_z v_x^2\partial_u\log(v_z^2)}{\omega^2-(k_z^2 v_z^2+k_x^2 v_x^2)}E'_x-\frac{G^{tt}}{G^{uu}}\left(\omega^2-(k_z^2 v_z^2+k_x^2 v_x^2)\right)E_z=0 \ .
\eear

We can now start solving these equations. There are two important cases that can be solved analytically. For $T=0$, we will find a zero sound like dispersion relation and for $T\neq 0$ we will find a diffusion dispersion relation. The strategy for both cases involves solving the equations when $\omega,k\to0$ with the near-horizon limit and then demanding that these two limits commute. We will begin with the $T=0$ case.

\subsection{$T=0$: zero sound mode}

First, we take the limit $u\to \infty$. We need the asymptotic behavior of the metric components,
\bear
\mathcal{F}= F_0 u^{\alpha_f}\left(1+\mathcal{O}\left(\frac{1}{u}\right)\right),\quad \mathcal{B}= B_0 u^{-\alpha_b}\left(1+\mathcal{O}\left(\frac{1}{u}\right)\right),\quad \phi= {\tilde \phi}_0-\alpha_{\phi}\log\left(\frac{u}{L_s} +\mathcal{O}\left(1\right)\right) \ .
\rc
\eear
To preserve Lorentz invariance for the $(t,x,y)$ components, we set $\alpha_b=\alpha_f$. We relax the conditions on \eqref{eq:simple} by setting $\mathcal{H}=H_0 {\rm e}^{-\phi}$, $\mathcal{Z}=Z_0 {\rm e}^{\phi/2}$. 
The series expansions below are valid if $\alpha_{\phi}<4$, $\alpha_f>-2$, and $\alpha_f-2\alpha_{\phi}>-2$, which are the cases of interest to us, {\emph{i.e.}}, for the pure AdS metric and for the metric appearin later in \eqref{eq:alt}.

The equations of motion for $E$ decouple in the asymptotic limit
\bear
E_x'' +\frac{4+\alpha_f}{2u}E'_x+\frac{\omega^2}{B_0 F_0^2 u^{\alpha_f}}E_x&=&0 \,\,\rc\rc
E_z'' +\frac{4+\alpha_f-2\alpha_{\phi}}{2u}E'_z+\frac{\omega^2}{B_0 F_0^2 u^{\alpha_f}}E_z&=&0 \ .
\eear
We require ingoing boundary conditions so the solution to these equations is
\bear
E_q = \frac{F_{q}}{u^{r_q/2}}H_{\frac{r_q}{2-\alpha_f}}^{(1)}\left(\frac{2u^{1-\frac{\alpha_f}{2}}\omega}{\sqrt{B_0}F_0(2-\alpha_f)}\right),\qquad r_x=\frac{2+\alpha_f}{2},\qquad r_z = \frac{2+\alpha_f-2\alpha_{\phi}}{2} \ ,
\qquad
\eear
where $F_q$ is an integration constant and $H_{r_q}^{(1)}$ is the Hankel function of the first kind. The $\omega\to 0$ limit of these are
\beq
E_q\approx F_q\left(\frac{1+i\cot\left(\frac{\pi r_q}{2-\alpha_f}\right)}{\Gamma \left(1+\frac{r_q}{2-\alpha_f}\right)}\frac{\omega^{\frac{r_q}{2-\alpha_f}}}{(\sqrt{B_0} F_0 \left(2-\alpha _f\right))^{\frac{r_q}{2-\alpha_f}}}- \frac{i \Gamma \left(\frac{r_q}{2-\alpha_f}\right)}{u^{r_q}\pi}\frac{(\sqrt{B_0} F_0 \left(2-\alpha_f\right))^{\frac{r_q}{2-\alpha_f}}}{\omega^{\frac{r_q}{2-\alpha_f}}}\right)\ .\label{eq:zerotemexp1}
\eeq

Second, we take the low-frequency limit of equations \eqref{eq:as}. The equations decouple and $a'_x$ and $a'_z$ are easily solved. From these, we obtain $E'_x$ and $E'_z$ and then integrate them
\beq
a''_q+\partial_u\log\left(\sqrt{g_{uu}|g_{tt}|}\frac{W}{\sqrt{W+d^2}}G^{uu}G^{qq} \right)a'_q =0, 
\qquad\quad
q=x,z \ .
\eeq
These are easily integrated and the gauge-invariant fields are
\bear
E_x\!\! &=&\!\! E_{x,0}+\int \limits_{0}^{u}d{\tilde u}\Bigg[
 \frac{C_x\omega }{{\tilde u}^2\sqrt{d^2+{\rm e}^{-\phi/2}\mathcal{H}Z_0^3L_s^{12}{\tilde u}^{-6}}\sqrt{\mathcal{B}}\mathcal{F}}-\frac{k_x}{\omega}\,\frac{{\rm e}^{-\frac{\phi}{2}}\mathcal{H} Z_0^3L_s^{12}\sqrt{\mathcal{B}}\left(C_xk_x+C_z k_z\right)}{{\tilde u}^8(d^2+{\rm e}^{-\phi/2}\mathcal{H}Z_0^3L_s^{12}{\tilde u}^{-6} )^{{3\over 2}}}\Bigg]
\rc
\!\!&\equiv&\!\! E_{x,0}+\omega C_x{\cal J}_x(u)-\frac{k_x(C_xk_x+C_zk_z)}{\omega} {\cal  I}(u)\,\rc
E_z\!\! &=&\!\! E_{z,0}+\int \limits_{0}^{u}{\tilde u} \Bigg[
\frac{C_z\omega \mathcal{H}}{{\tilde u}^2\sqrt{d^2+{\rm e}^{-\phi/2}\mathcal{H}Z_0^3L_s^{12}{\tilde u}^{-6}}\sqrt{\mathcal{B}}\mathcal{F}}-\frac{k_z}{\omega}\,\frac{{\rm e}^{-\frac{\phi}{2}}\mathcal{H} Z_0^3L_s^{12}\sqrt{\mathcal{B}}\left(C_xk_x+C_z k_z\right) }{{\tilde u}^8 (d^2+{\rm e}^{-\phi/2}\mathcal{H}Z_0^3L_s^{12}{\tilde u}^{-6} )^{{3\over 2}}}
\Bigg]\rc
\!\!&\equiv&\!\! E_{z,0}+\omega C_z{\cal J}_z(u)-\frac{k_z(C_xk_x+C_zk_z)}{\omega}{\cal I}(u) \ ,
\eear
where, in the second step, we have defined the integrals ${\cal J}_x(u)$,  ${\cal J}_z(u)$, and 
${\cal I}(u)$.  In the above expressions, $C_x$, $C_z$, $E_{x,0}$, and $E_{z,0}$ are integration constants. The next step is to approximate these expressions at the near-horizon limit
\bear
E_x\!\!\!\!\! &=&\!\!\!\!\! E_{x,0}+\omega \,C_x{\hat J}_x-\frac{k_x(C_xk_x+C_zk_z)}{\omega}{\hat I}-\omega\, C_x\, \frac{2}{\sqrt{B_0}dF_0\left(2+\alpha_f\right)} u^{-\frac{\alpha_f}{2}-1}\,\,\rc
E_z\!\!\!\!\! &=&\!\!\!\!\! E_{z,0}+\omega\, C_z{\hat J}_z-\frac{k_z(C_xk_x+C_zk_z)}{\omega}{\hat I}-\omega\, C_z\,\frac{2 {\rm e}^{-{\tilde \phi}_0} H_0 }{\sqrt{B_0} d F_0 \left(2-2 \alpha_{\phi}+\alpha_f\right)}\frac{u^{\alpha_{\phi }-\frac{\alpha _f}{2}-1}}{L_s^{\alpha_{\phi}}} \ ,
\eear
where
\beq
{\hat J}_x\,\equiv \,{\cal J}_x(u=\infty)\,\,,\qquad\qquad
{\hat J}_z\,\equiv \,{\cal J}_z(u=\infty)\,\,,\qquad\qquad
{\hat I}\,\equiv \,{\cal I}(u=\infty)\,\,.
\eeq
We now match these expansions with the ones in \eqref{eq:zerotemexp1}. We first solve the $F_x$ and $F_z$ coefficients and then solve for $E_{x,0}$ and $E_{z,0}$, which will give us two linear equations in terms of $C_x$ and $C_z$. Imposing the Dirichlet boundary conditions ($E_{q,0}=0$), the only way to obtain a non-trivial solution is to require singularity of the linear equation, which will give us the dispersion relation. After a few straightforward steps, we get
\bear
\left(\begin{array}{cc} E_{x,0} \\ E_{z,0} \end{array}\right)&=&-\frac{1}{\omega}\times\nonumber\\
&&\!\!\!\!\!\!\!\!\!\!\!\!\times\left(\begin{array}{cc} -{\hat I}k_x^2 +{\hat J}_x\omega^2-\mu_x \omega^{\frac{6-\alpha_f}{2-\alpha_f}}& -{\hat I}k_xk_z \\
 -{\hat I}k_xk_z & -{\hat I}k_z^2+{\hat J}_z\omega^2- \mu_z\omega^{\frac{6-\alpha_f-2\alpha_{\phi}}{2-\alpha_f}} \end{array}\right) \label{disp-matrix}\left(\begin{array}{c} C_x \\ C_z \end{array}\right)\label{eq:zero-matrix}  ,\qquad\qquad
\eear
where the coefficients are
\bear
\mu_x &=& -\frac{\pi\left(\tan\left(\frac{\pi\alpha_f}{2-\alpha_f}\right)+i\right)  \left(\frac{1}{\sqrt{B_0} F_0 \left(2-\alpha_f\right)}\right)^{\frac{4}{2-\alpha_f}}}{d\Gamma\left(\frac{6-\alpha_f}{4-2 \alpha_f}\right)^2}\,\rc
\mu_z &=& \frac{\pi H_0{\rm e}^{-{\tilde \phi}_0} \left(\tan \left(\frac{\pi \left(\alpha_{\phi}-\alpha_f\right)}{2-\alpha_f}\right)-i\right)\left(\frac{1}{\sqrt{B_0}F_0\left(2-\alpha_f\right)}\right)^{\frac{4-2\alpha_{\phi}}{2-\alpha_f}}}{dL_s^{\alpha_{\phi}}\Gamma \left(\frac{6-\alpha_f-2\alpha_{\phi}}{4-2\alpha_f}\right)^2} \ .
\eear
The condition for the determinant to be zero, {\emph i.e.}, the dispersion relation, is given by the equation 
\beq
\left(\frac{{\hat J}_z}{{\hat I}}\omega^2-k_z^2-\omega^{\frac{6-\alpha_f-2\alpha_{\phi}}{2-\alpha_f}} \frac{\mu_z}{{\hat I}}\right)\left(\frac{{\hat J}_x}{{\hat I}}\omega^2-k_x^2- \frac{\mu_x}{{\hat I}} \omega^{\frac{6-\alpha_f}{2-\alpha_f}}\right)=k_x^2k_z^2\ .\label{eq:0sound}
\eeq

The 1st order solution to this equation gives us the zero sound mode
\beq
\omega^2 = k_x^2 \frac{{\hat I}}{{\hat J}_x}+k_z^2 \frac{{\hat I}}{{\hat J}_z}=k^2\left(\frac{{\hat I}}{{\hat J}_x}\cos^2\varphi+\frac{{\hat I}}{{\hat J}_z}\sin^2\varphi\right)\equiv (c_x^2\cos\varphi^2+c_z^2\sin\varphi^2)k^2\equiv k^2 c_q^2\ .\label{eq:0soundspeed}
\eeq
The next order terms gives us the damping of the mode. By defining $\omega = c_q k +\delta\omega$, we can extract
\beq
\delta\omega=k^{\frac{4}{2-\alpha_f}}\frac{c_q^{\frac{2\alpha_f}{2-\alpha_f}}{\hat I}}{2}\left(\frac{\mu_x}{{\hat J}_x^2}\cos^2\varphi+(kc_q)^{\frac{-2\alpha_{\phi}}{2-\alpha_f}}\frac{\mu_z}{{\hat J}_z^2}\sin^2\varphi\right) \ .\label{eq:0soundcorrection}
\eeq


\subsection{$T\neq 0$: diffusion mode}
This time we set $T\neq 0$ and take the near-horizon limit $u\to u_H$ and the low-frequency limit $\omega\sim k^2\to 0$. Note that we are implicitly expecting to find a diffusive solution, {\emph i.e.}, $\omega=-i Dk^2$.

We first take the near horizon limit of the equations of motion 
\bear
\nonumber
E_x'' + \left(\frac{1}{u-u_H}+b_x\right)E_x' + \left(\frac{c_x}{(u-u_H)^2}+\frac{d_x}{u-u_H}\right)E_x+ f_xE_z'&=&0\\
E_z'' + \left(\frac{1}{u-u_H}+b_z\right)E_z' + \left(\frac{c_z}{(u-u_H)^2}+\frac{d_z}{u-u_H}\right)E_z+ f_zE_x'&=&0 \ ,
\eear
where the coefficients are
\bear
b_x &=&\frac{\mathcal{B}'\left(u_H\right)}{2\mathcal{B}\left(u_H\right)}+\frac{2}{u_H}-\frac{H_0 Z_0^3L_s^{12}(\frac{3}{4}\phi'\left(u_H\right)+3 u_H^{-1}-\frac{k_x^2}{\omega^2}\mathcal{B}(u_H)\mathcal{F}'(u_H))}{(d^2 u_H^6 e^{\frac{3 \phi \left(u_H\right)}{2}}+H_0 Z_0^3L_s^{12})}+\frac{\mathcal{F}''\left(u_H\right)}{2\mathcal{F}'\left(u_H\right)}\,\,\rc\rc
b_z &=& \frac{\mathcal{B}'\left(u_H\right)}{2\mathcal{B}\left(u_H\right)}+\frac{2}{u_H}+\phi'(u_H)-\frac{H_0 Z_0^3L_s^{12} (3u_H \phi '\left(u_H\right)+12)}{4u_H(d^2 u_H^6 e^{\frac{3 \phi \left(u_H\right)}{2}}+H_0 Z_0^3L_s^{12} )}+\frac{ \mathcal{F}''\left(u_H\right)}{2\mathcal{F}'\left(u_H\right)}\rc\rc
&&\quad\quad\quad+\frac{Z_0^3L_s^{12} k_z^2 \mathcal{B}\left(u_H\right) e^{\phi \left(u_H\right)} \mathcal{F}'\left(u_H\right)}{\omega ^2 \left(d^2 u_H^6 e^{\frac{3\phi\left(u_H\right)}{2}}+H_0 Z_0^3L_s^{12}\right)}\,\rc\rc
f_x &=& \frac{Z_0^3 L_s^{12}k_x k_z \mathcal{B}\left(u_H\right) e^{\frac{\phi \left(u_H\right)}{2}} \mathcal{F}'\left(u_H\right)}{\omega^2\left(d^2 u_H^6 e^{\phi\left(u_H\right)}+H_0 Z_0^3L_s^{12} e^{-\frac{1}{2} \phi \left(u_H\right)}\right)} = \frac{e^{\phi \left(u_H\right)}}{H_0}f_z\,\,\rc\rc
d_x &=& d_z = -\frac{\omega ^2 \left(\mathcal{B}'\left(u_H\right) \mathcal{F}'\left(u_H\right)+\mathcal{B}\left(u_H\right) \mathcal{F}''\left(u_H\right)\right)}{\mathcal{B}\left(u_H\right){}^2 \mathcal{F}'\left(u_H\right)^3}-\frac{L_s^{12}Z_0^3 \left(k_z^2 e^{\phi \left(u_H\right)}+H_0 k_x^2\right)}{\mathcal{F}'\left(u_H\right) \left(d^2 u_H^6 e^{\frac{3 \phi \left(u_H\right)}{2}}+H_0 Z_0^3L_s^{12}\right)}\,\rc\rc
c_x&=&c_z = \frac{\omega^2}{\mathcal{B}(u_H)\mathcal{F}'(u_H)^2} \ .
\eear

We solve these equations using the Frobenius series, {\emph{i.e.}}, $E_q = F_q(u-u_H)^{\alpha_q}(1+\beta_q(u-u_H)+\ldots)$, where $\alpha_q$, $F_q$ and $\beta_q$ are all coefficients that might depend on $\omega$ and $k$. With the expansion, we can solve for $\alpha_q$'s and $\beta_q$'s
\bear
\alpha_q &=& -i\sqrt{c_q}=-i\frac{\omega}{\sqrt{\mathcal{B}(u_H)}|\mathcal{F}'(u_H)|} \equiv \alpha\,\rc
\beta_{x,z}&=& -\frac{d_{x,z}F_{x,z}+\alpha(b_{x,z}F_{x,z}+f_{x,z}F_{z,x})}{(1+2\alpha)F_{x,z}} \ ,
\eear
where we have chosen an explicit sign for $\alpha_q$ in order to have an infalling solution. In the expression for $\beta_q$, one either chooses the first indices or the second indices for all the terms. Taking the low-frequency limit with $\omega\sim k^2\sim \epsilon^2$, $\beta$'s take the value
\beq
\frac{F_q\beta_q}{k_q}=-\frac{iZ_0^3L_s^{12}e^{\phi\left(u_H\right)} \sqrt{\mathcal{B}\left(u_H\right)}\left(F_xH_0 e^{-\phi\left(u_H\right)}k_x+F_zk_z\right)}{\omega\left(d^2u_H^6e^{\frac{3\phi\left(u_H\right)}{2}}+H_0Z_0^3L_s^{12}\right)}+\mathcal{O}(\varepsilon^2)\ .
\eeq

For the other order, we once again first solve for $a'_x$ and $a'_z$ and then write down the $E'_q$. The solution reads
\beq
a'_q = C_q\left(\sqrt{g_{uu}|g_{tt}|}\frac{W}{\sqrt{W+d^2}}G^{uu}G^{qq} \right)^{-1} \ ,\quad q=x,z \ ,
\eeq
which we then plug into the expression for $E'$'s
\bear
E_x' &=& \frac{\omega^2a'_x - (k_x^2v_x^2 a'_x +k_zk_x a'_z v_z^2)}{\omega} \label{Ex-derivT}
\,\,\rc
E_z' &=& \frac{\omega^2a'_z - (k_z^2 v_z^2 a'_z+k_xk_z a'_x v_x^2)}{\omega}\ . \label{Ez-derivT}
\eear
To respect our low-frequency expansion, we can neglect the $\omega^2$ terms as $a'_x$ and $a'_z$ should be of the same order.

The integrated expressions for $E$'s then read
\bear
E_q\!\! &=&\!\! E_{q,0} - \int\limits_{0}^{u}{\rm d}{\tilde u}\,\frac{k_q}{\omega}\,\frac{e^{-\frac{\phi}{2}}\mathcal{H} Z_0^3L_s^{12}\sqrt{\mathcal{B}({\tilde u})} \left(C_xk_x+C_z k_z\right)}{{\tilde u}^8  \left(d^2+{H_0 Z_0^3L_s^{12} e^{-3\phi ({\tilde u})/2}{\tilde u}^{-6}}\right)^{3/2}}\label{eq:Exd-diff}\,\,\rc
\!\!&=&\!\! E_{q,0}-\frac{k_q(C_x k_x+C_z k_z)}{\omega}\mathcal{I}(u)\,\,\rc
\!\!&\approx&\!\! E_{q,0}-\frac{k_q(C_x k_x+C_z k_z)}{\omega}\left(\mathcal{I}(u_H)+\frac{e^{-\frac{\phi \left(u_H\right)}{2}} \mathcal{H}(u_H)Z_0^3L_s^{12} \sqrt{\mathcal{B}\left(u_H\right)} }{u_H^8 \left(d^2+{H_0 Z_0^3L_s^{12} e^{-3\phi \left(u_H\right)/2}u_H^{-6}}\right)^{3/2}}(u-u_H)\right)\ ,\qquad\qquad
\eear
where we have also done a near-horizon expansion.

We must now match our two solutions. First, we set $(u-u_H)^{\alpha}=E_{\rm nh}$ and $C_q\propto F_q$, then match the two terms in the expansions. We get the relation
\beq
\left(\begin{array}{c}E_{x,0}\\E_{z,0}\end{array}\right)= \frac{-i}{\omega u_H^2\sqrt{d^2+{\rm e}^{-3\phi(u_H)/2}H_0Z_0^3L_s^{12}u_H^{-6}}}\left(\begin{array}{cc}\omega+ik_x^2 D_x&ik_xk_zD_x\\ik_xk_zD_x&\mathcal{H}(u_H)\left(\omega+ik_z^2\frac{ D_x}{\mathcal{H}(u_H)}\right)\end{array}\right) \left(\begin{array}{c}C_{x}\\C_{z}\end{array}\right),\label{matrix-diffusion}
\eeq
where
\beq
D_x = \mathcal{I}(u_H)u_H^2\sqrt{d^2+{\rm e}^{-3\phi(u_H)/2}Z_0^3H_0L_s^{12}u_H^{-6}} \ .
\eeq
A non-trivial solution to Dirichlet boundary conditions is provided only when the matrix is singular, {\emph i.e.}, when
\beq
\omega = -i (k_x^2 + k_z^2/\mathcal{H}(u_H)) D_x\ ,\label{eq:diffusion} 
\eeq
which corresponds to a diffusion mode.


\subsection{Two-point functions}
We now move on to compute two-point functions of this system both in zero temperature and finite temperature. From these two-point functions we can extract the conductivity with which we can do a non-trivial consistency check through the Einstein relation.

Including all prefactors in \eqref{eq:lagrangian2order}, we get the 2nd order Lagrangian
\beq
\mathcal{L}=-T_7L_s^3\sqrt{|g_{tt}|g_{uu}}\frac{W}{\sqrt{W+d^2}}\left(\sum\limits_{A<B}^{}\frac{G^{AA}G^{BB}(f_{AB})^2}{2}\right)\ .
\eeq
First of all, we drop all the fields in directions other than $t$, $z$, or $x$. To use gauge invariant quantities, we write the bracketed sum first as
\bear
&&\sum\limits_{A<B}^{} G^{AA}G^{BB}(f_{AB})^2 = G^{uu}\left(G^{tt}(a'_t)^2+G^{xx}(a'_x)^2+G^{zz}(a'_z)^2\right)\rc
&&\quad\quad\quad\quad\quad+ G^{tt}G^{xx}E_x^2 +G^{tt}G^{zz}E_z^2 +\frac{1}{\omega^2}G^{xx}G^{zz}(E_x k_z-E_zk_x)^2\ , 
\label{eq:dotted}
\eear
where we already performed a Fourier transform. All the multiples of fields are to be interpreted as
\beq
E_q E_{q'}= E_q^*(k,\omega)E_{q'}(k,\omega)\ .
\eeq
The term with derivatives with respect to $u$ can then be written as
\beq
\!\!\!\frac{G^{xx}(G^{zz}k_z^2+G^{tt}\omega^2)(E_x')^2 +G^{zz}(G^{xx}k_x^2+G^{tt}\omega^2) (E_z')^2  
- \left(G^{xx}G^{zz}k_xk_z \right)(E_z'E_x'+E_x'E_z')}{\omega^2(\omega^2 G^{tt}+k_z^2 G^{zz}+k_x^2 G^{xx})/G^{uu}}\ .
\eeq

We now do a partial integration with respect to $u$ for the terms which include $E_q'$ terms. The strategy is to integrate the complex conjugated field and then differentiate the rest of the expression. Making use of equations of motion, it turns out that all the remaining bulk integrals cancel so we are left with only a surface integral.
\bear
&&S_{\rm{on-shell}}^{(2)}= \frac{T_7\Omega_3L_s^3}{2}\int d^4{k}\sqrt{|g_{tt}|g_{uu}}\frac{WG^{uu}}{\sqrt{W+d^2}}\rc\rc
&&\!\!\!\!\!\!\!\!\!\!\!\!\left.\times\frac{G^{xx}(G^{zz}k_z^2+G^{tt}\omega^2)E_xE_x' +G^{zz}(G^{xx}k_x^2+G^{tt}\omega^2) E_zE_z' - \left(G^{xx}G^{zz}k_xk_z \right)(E_zE_x'+E_xE_z')}{\omega^2(\omega^2 G^{tt}+k_z^2 G^{zz}+k_x^2 G^{xx})}\right|_{u\to 0}\!\!\!\!\!\! \ .\rc\rc
\eear
As a next step, we consider a low-energy limit, yielding
\beq
S_{\rm{on-shell}}^{(2)}= \frac{T_7\Omega_3L_s^3}{2}\int d^4{k} \frac{1}{\omega}(E_{x,0}(k)^*C_x(k)+E_{z,0}(k)^*C_z(k))\ .
\eeq
The final steps involve expressing $C_q$'s in terms of the boundary values and taking functional derivatives with respect to the boundary values.


\subsubsection{$T=0$}
We invert equation \eqref{eq:zero-matrix} and get
\bear
\left(\begin{array}{c}C_x \\ C_z \end{array}\right) &=& \frac{\omega}{\left({\hat I}k_z^2-{\hat J}_z\omega^2 +\mu_z\omega^{\frac{6-\alpha_f-2\alpha_{\phi}}{2-\alpha_f}}\right)\left({\hat I}k_x^2-{\hat J}_x\omega^2+\mu_x\omega^{\frac{6-\alpha_f}{2-\alpha_f}}\right)-{\hat I} k_x^2k_z^2}\\
&&\times\left(\begin{array}{cc} {\hat I}k_z^2-{\hat J}_z\omega^2 +\mu_z\omega^{\frac{6-\alpha_f-2\alpha_{\phi}}{2-\alpha_f}}& -{\hat I}k_xk_z \\ -{\hat I}k_xk_z & {\hat I}k_x^2-{\hat J}_x\omega^2+\mu_x\omega^{\frac{6-\alpha_f}{2-\alpha_f}}  \end{array}\right)\left(\begin{array}{c}E_{x,0} \\ E_{z,0} \end{array}\right)\nonumber \ .
\eear

The on-shell action, in terms of the boundary values, now takes the form
\bear
&&\!\!\!\!\!\!\!\!\!\!\!S_{\rm{on-shell}}^{(2)}= \frac{T_7\Omega_3L_s^3}{2}\!\!\int \!\!d^4{k}\,\frac{1}{\left({\hat I}k_z^2-{\hat J}_z\omega^2 +\mu_z\omega^{\frac{6-\alpha_f-2\alpha_{\phi}}{2-\alpha_f}}\right)\left({\hat I}k_x^2-{\hat J}_x\omega^2+\mu_x\omega^{\frac{6-\alpha_f}{2-\alpha_f}}\right)-{\hat I}^2 k_x^2k_z^2}\rc\rc
&&\!\!\!\!\times\left(\left(\mu_z \omega^{\frac{6-\alpha_f-2\alpha_{\phi}}{2-\alpha_f}}-{\hat J}_z\omega^2+{\hat I}k_z^2\right)E_{x,0}^2+\left(\mu_x\omega^{\frac{6-\alpha_f}{2-\alpha_f}}-{\hat J}_x\omega^2+{\hat I}k_x^2\right)E_{z,0}^2\right.\nonumber\\
&&\quad\quad\quad\quad\quad\quad\quad\quad\quad\quad\quad\quad\quad\quad\quad\quad\quad\quad-{\hat I}k_xk_z(E_{x,0}E_{z,0}+E_{z,0}E_{x,0})\bigg) \ .
\eear
Now we will simply take functional derivatives of the on-shell action to obtain the current-current correlators. Bear in mind that $\frac{\partial }{\partial a_{\mu}(k)}=\frac{\partial E_i(k)}{\partial a_{\mu}(k)}\frac{\partial }{\partial E_i(k)}$. First, the $tt$ correlator is
\bear
&&\langle J_t(-k) J_t(k)\rangle = \frac{\delta }{\delta a_t^*(k)}\frac{\delta }{\delta a_t(k)}S^{(2)}\rc\rc
&=& \frac{T_7\Omega_3L_s^3\left(\mu_x\omega^{\frac{6-\alpha_f}{2-\alpha_f}}k_z^2+\mu_z\omega^{\frac{6-\alpha_f-2\alpha_{\phi}}{2-\alpha_f}}k_x^2-\omega^2\left(k_x^2{\hat J}_z+k_z^2{\hat J}_x \right) \right)}{\left({\hat I}k_z^2-{\hat J}_z\omega^2 +\mu_z\omega^{\frac{6-\alpha_f-2\alpha_{\phi}}{2-\alpha_f}}\right)\left({\hat I}k_x^2-{\hat J}_x\omega^2+\mu_x\omega^{\frac{6-\alpha_f}{2-\alpha_f}}\right)-{\hat I}^2 k_x^2k_z^2}\ .
\eear

Second, we consider direction $q=\cos\chi\, {\hat x}+ \sin\chi \, {\hat z}$, {\emph i.e.} the current $J_q = J_x \cos\chi+J_z \sin\chi$. In addition, for a more condensed notation, we use the notation $k_x=k\cos\varphi$ and $k_z=k\sin\varphi$. The current two-point function becomes
\bear
&&\!\!\!\!\!\!\!\!\!\!\!\!\!\!\!\!\!\!\langle J_q(-k) J_q(k)\rangle = \langle J_x(-k) J_x(k)\rangle\cos^2\chi+ \langle J_z(-k) J_z(k)\rangle\sin^2\chi+\frac{\langle J_x(-k) J_z(k)+J_z(-k) J_x(k)\rangle}{2}\sin 2\chi \rc\rc
&&\!\!\!\!\!\!\!\!\!\!= T_7\omega^2\Omega_3L_s^3\frac{\mu_x\omega^{\frac{6-\alpha_f}{2-\alpha_f}}\sin^2\chi+\mu_z\omega^{\frac{6-\alpha_f-2\alpha_{\phi}}{2-\alpha_f}}\cos^2\chi-\omega^2\left({\hat J}_z\cos^2\chi+{\hat J}_x\sin^2\chi \right)+{\hat I}k^2\sin^2(\varphi-\chi) }{\left({\hat I}k_z^2-{\hat J}_z\omega^2 +\mu_z\omega^{\frac{6-\alpha_f-2\alpha_{\phi}}{2-\alpha_f}}\right)\left({\hat I}k_x^2-{\hat J}_x\omega^2+\mu_x\omega^{\frac{6-\alpha_f}{2-\alpha_f}}\right)-{\hat I}^2 k_x^2k_z^2} \ .\rc
\eear
On the other hand, two perpendicular directions with $q_{\perp} = \sin\chi\, {\hat x}- \cos\chi \, {\hat z}$ have the two-point function
\bear
&&\langle J_q(-k)J_{q_{\perp}}(k)\rangle\rc\rc
&&\!\!\!\!\!\!\!\!\!\!= \frac{T_7\omega^2\Omega_3L_s^3\left(\sin\chi\cos\chi\left[\mu_z\omega^{\frac{6-\alpha_f-2\alpha_{\phi}}{2-\alpha_f}}-\mu_x\omega^{\frac{6-\alpha_f}{2-\alpha_f}}-\omega^2\left({\hat J}_z-{\hat J}_x\right)\right]+{\hat I}\frac{k^2\sin(2(\varphi-\chi))}{2}\right)}{\left({\hat I}k_z^2-{\hat J}_z\omega^2 +\mu_z\omega^{\frac{6-\alpha_f-2\alpha_{\phi}}{2-\alpha_f}}\right)\left({\hat I}k_x^2-{\hat J}_x\omega^2+\mu_x\omega^{\frac{6-\alpha_f}{2-\alpha_f}}\right)-{\hat I}^2 k_x^2k_z^2} \ .\qquad\qquad
\eear

The conductivity tensor can be computed with the relation
\beq
\sigma_{ij}(\omega)=\frac{1}{i\omega}\langle J_i(-\omega,{\vec k}=0)J_j(\omega,{\vec k}=0)\rangle \ .
\eeq
In the low-frequency limit, the longitudinal and Hall conductivities are
\beq
\sigma_{qq}(\omega)=i\frac{T_7\Omega_3L_s^3\left({\hat J}_z\cos^2\chi+{\hat J}_x\sin^2\chi\right)}{{\hat J}_z{\hat J}_x}\frac{1}{\omega}\ ,
\eeq
and
\beq
\sigma_{qq_{\perp}}(\omega)=i\frac{T_7\Omega_3L_s^3\sin\chi\cos\chi\left({\hat J}_z-{\hat J}_x\right)}{{\hat J}_z{\hat J}_x}\frac{1}{\omega} \ .
\eeq
The $\frac{i}{\omega}$ singularity resembles Drude conductivity and implies a delta peak for the real part of the conductivity at $\omega=0$.

\subsubsection{$T\neq 0$}

Inverting the matrix in \eqref{matrix-diffusion} gives
\beq
\left(\begin{array}{c}C_{x}\\C_{z}\end{array}\right)= i\frac{u_H^2\sqrt{d^2+{\rm e}^{-3\phi(u_H)/2}H_0Z_0^3L_s^{12}u_H^{-6}}}{\mathcal{H}(u_H)	(\omega+i(k_x^2+\frac{k_z^2}{\mathcal{H}(u_H)})D_x)}\left(\begin{array}{cc}\mathcal{H}(u_H)(\omega+ik_z^2\frac{ D_x}{\mathcal{H}(u_H)})&-ik_xk_zD_x\\-ik_xk_zD_x&\omega+ik_x^2 D_x\end{array}\right) \left(\begin{array}{c}E_{x,0}\\E_{z,0}\end{array}\right) \ ,
\eeq
which we use to express the on-shell action in terms of field boundary values
\bear
&&S_{\rm{on-shell}}^{(2)}=\frac{iT_7\Omega_3L_s^{3}\sqrt{d^2+{\rm e}^{-3\phi(u_H)/2}H_0Z_0^3L_s^{12}u_H^{-6}}u_H^2}{2}\rc\rc
&&\!\!\!\!\!\!\!\!\times\!\!\int\!\! {\rm d}^4{k} \frac{\mathcal{H}(u_H)(\omega+i \frac{D_x}{\mathcal{H}(u_H)} k_z^2)E_{x,0}^2+ (\omega + i D_x k_x^2)E_{z,0}^2-ik_xk_zD_x(E_{z,0}E_{x,0}+E_{x,0}E_{z,0})}{\omega\mathcal{H}(u_H)(\omega+i(k_x^2+\frac{k_z^2}{\mathcal{H}(u_H)})D_x)}\,\,.\rc
\eear

The two-point functions can then be easily computed. We list the same components as above
\bear
&&\langle J_t(-k)J_t(k)\rangle\! =\! \frac{iT_7\Omega_3L_s^{3}\sqrt{d^2+{\rm e}^{-3\phi(u_H)/2}H_0Z_0^3L_s^{12}u_H^{-6}} u_H^2\left(\mathcal{H}(u_H)k_x^2+k_z^2\right)}{\mathcal{H}(u_H)(\omega+i(k_x^2+\frac{k_z^2}{\mathcal{H}(u_H)})D_x)}\,\,\rc\rc
&&\langle J_q(-k)J_q(k)\rangle\!\!\! \rc\rc
&& =\frac{iT_7\Omega_3L_s^{3}\omega\sqrt{d^2+{\rm e}^{-3\phi(u_H)/2}H_0Z_0^3L_s^{12}u_H^{-6}}u_H^2 \left(\omega(\mathcal{H}(u_H)\cos^2\chi+\sin^2\chi)+ ik^2D_x\sin^2(\varphi-\chi)\right)}{\mathcal{H}(u_H)(\omega+i(k_x^2+\frac{k_z^2}{\mathcal{H}(u_H)})D_x)}\,\,\rc\rc
&&\langle J_q(-k)J_{q_{\perp}}(k)\rangle \rc\rc
&&= \frac{iT_7\Omega_3L_s^{3}\omega\sqrt{d^2+{\rm e}^{-3\phi(u_H)/2}H_0Z_0^3L_s^{12}u_H^{-6}}u_H^2\left(\omega\cos\chi\sin\chi(\mathcal{H}(u_H)-1)+ i\frac{k^2D_x\sin 2(\varphi-\chi)}{2}\right)}{\mathcal{H}(u_H)(\omega+i(k_x^2+\frac{k_z^2}{\mathcal{H}(u_H)})D_x)}\,\,.
\eear
The corresponding DC conductivities are
\bear
\sigma_{qq} &=& T_7\Omega_3L_s^{3}\sqrt{d^2+{\rm e}^{-2\phi(u_H)}\mathcal{H}(u_H)\mathcal{Z}(u_H)^3L_s^{12}u_H^{-6}} u_H^2\left(\cos^2\chi+\frac{\sin^2\chi}{\mathcal{H}(u_H)}\right)\,\, \rc\rc
\sigma_{qq_{\perp}} &=& T_7\Omega_3L_s^{3}\sqrt{d^2+{\rm e}^{-2\phi(u_H)}\mathcal{H}(u_H)\mathcal{Z}(u_H)^3L_s^{12}u_H^{-6}}u_H^2 \left(1-\frac{1}{\mathcal{H}(u_H)}\right)\cos\chi\sin\chi\ .
\eear
We see that these only depend on the near-horizon physics.

\subsubsection{Einstein relation}
Einstein relation relates conductivity to susceptibility and diffusion in a non-trivial manner,
\beq
\sigma = D \chi_c.
\eeq
The validity of this relation has previously been checked and verified in many other holographic settings, starting with \cite{Mas:2008qs}. Susceptibility is computed with
\beq
\chi_c^{-1} = \left(\frac{\partial \mu}{\partial \rho}\right)_T =\frac{L_s^{9}}{T_7\Omega_3} \int\limits_{0}^{u_H}{\rm d}u \frac{{\rm e}^{-2\phi}\sqrt{\mathcal{B}}\mathcal{H}\mathcal{Z}^3}{u^8\left(d^2+{\rm e}^{-2\phi}\mathcal{H}\mathcal{Z}^3L_s^{12}u^{-6}\right)^{3/2}}=\frac{\mathcal{I}(u_H)}{T_7\Omega_3L_s^{3}}\ .
\eeq
A simple algebraic exercise shows that the Einstein relation is indeed satisfied.


\section{Fixed-point Lifshitz metric}\label{sec:lifshitz}

No analytical non-trivial solutions for the system discussed in Sec.~\ref{setup} are known. We wish to compare our analytical low-energy expressions to numerical ones. If we relax the regularity conditions near the boundary, we can find a closed form fixed-point Lifshitz-like solution, originally discovered in \cite{Azeyanagi:2009pr}. The solution is
\bear
&\alpha=\frac{\sqrt{6}a\mathcal{Z}_0^{5/2}}{\sqrt{\mathcal{H}_0}}L_s^5,\quad{\rm e}^{\phi}={\rm e}^{\phi_0}\left(\frac{L_s}{u}\right)^{4/7}=\sqrt{\frac{8}{3a^2L_s^2}}\sqrt{\frac{\mathcal{H}_0}{\mathcal{Z}_0}}\left(\frac{L_s}{u}\right)^{4/7}\!\!\!\!,\quad \mathcal{B}=\frac{33\mathcal{B}_0\mathcal{Z}_0}{49}\left(\frac{L_s}{u}\right)^{2/7}\!\!\!\!\,\, \rc\rc
&\mathcal{F}=\frac{49}{33\mathcal{Z}_0}\left(\frac{u}{L_s}\right)^{2/7}\left(1-\frac{u^{22/7}}{u_H^{22/7}}\right), \quad \mathcal{H}=\mathcal{H}_0\left(\frac{u}{L_s}\right)^{4/7},\quad \mathcal{Z}=\mathcal{Z}_0\left(\frac{L_s}{u}\right)^{2/7}  \label{eq:alt} \ ,
\eear
where $a$ is the axion parameter that determines the strength of the anisotropy and $\mathcal{B}_0$, $\mathcal{H}_0$, $\mathcal{Z}_0$ are free dimensionless parameters. Without losing generality, we can set them to unity. To obtain the original form of the fixed point metric in \cite{Azeyanagi:2009pr} from our solution above, we need to make a few modifications. By setting and scaling
\bear
u=\frac{L_s^8}{R_s^7}r^{-7/6}\ , \quad\{t,x,y\}\to \sqrt{\frac{11}{12}}\frac{L_s^7}{R_s^6}\{{\tilde t},{\tilde x},{\tilde y}\}\ ,\quad z\to\sqrt{\frac{11}{12}}\frac{L_s^5}{R_s^4}{\tilde w}\ ,\quad a= \sqrt{\frac{12}{11}}\frac{R_s^4}{L_s^5}\beta\,\,,\label{eq:alttomt} 
\eear
the constant $L_s$ drops from the expressions and we obtain the string frame metric
\bear
&&ds^2 = {\tilde R}_s^2\left(r^{7/3}(-f(r)d{\tilde t}^2+d{\tilde x}^2+d{\tilde y}^2)+r^{5/3}d{\tilde w}^2+\frac{dr^2}{r^{5/3}f(r)}\right)+R_s^2r^{1/3}d\Omega_{S^5}^2\rc
&&f(r) = 1-\left(\frac{r_H}{r}\right)^{11/3},\quad {\rm e}^{\phi}=\frac{\sqrt{22}}{3\beta}r^{2/3}, \quad {\tilde R}_s^2 = \frac{11}{12}R_s^2\ , \quad \alpha = \sqrt{\frac{72}{11}}R_s^4\beta\ .\label{eq:altmet}
\eear
In the Einstein frame $ds^2_E = {\rm e}^{-\phi/2}ds^2$ and with $r_H=0$, we see that the metric
\beq
ds_E^2 ={\tilde R}_E^2\left(r^2(-d{\tilde t}^2+d{\tilde x}^2+d{\tilde y}^2)+r^{4/3}d{\tilde w}^2+\frac{dr^2}{r^{2}}\right)+R_E^2d\Omega_{S^5}^2 \ ,
\eeq
exhibits a Lifshitz-like scaling with $\{{\tilde t},{\tilde x},{\tilde y}\}\to k\{{\tilde t},{\tilde x},{\tilde y}\}$, ${\tilde w}\to k^{2/3}{\tilde w}$, and $r\to k^{-1}r$. In the metric, $R_E^2 = \frac{3\beta}{\sqrt{22}}R_s^2$. 
Notice that one cannot consider the isotropic limit in this solution due to double scaling limit.

This solution was originally obtained by considering the one-form and the RR five-form to be sourced by $N$ D3- and $k$ D7-branes s.t. $\alpha=\frac{(2\pi)^4N}{Vol(S^5)}$, $\beta = \frac{k}{L_w}$, where $L_w$ is the period of the ${\tilde w}$ coordinate. From low-energy flat space perspective, the branes are extended along the following directions
\beq
\begin{array}{r|cccc|c|ccccc}
\mathcal{M}_4\times S^1\times X_5 & {\tilde t} & {\tilde x} & {\tilde y} & r & {\tilde w} & s_1 & s_2 & s_3 & s_4 & s_5\\
\hline 
N \, {\rm D3} & \times & \times &\times & & \times & & & & & \\
k \, {\rm D7} & \times & \times &\times & & & \times & \times &\times &\times &\times 
\end{array} \ .
\eeq

We will specialize our above solutions and computations to this special background metric. We will scale out thermal factors by redefining the radial coordinate and setting the horizon to reside at 1. The rescaled quantities we will decorate with hats as follows:
\bear
&&\omega = {\hat \omega}u_H^{-6/7}L_s^{-1/7},\quad k_x = {\hat k_x}u_H^{-6/7}L_s^{-1/7},\quad k_z = {\hat k_z}u_H^{-4/7}L_s^{-3/7} 
\,\,\rc\rc
&&E_x = {\hat E}_x u_H^{-2}L_s^{2},\quad
E_z = {\hat E}_z u_H^{-12/7}L_s^{12/7}\ .
\eear
The only dimensionless parameter we thus have is the rescaled charge density ${\hat d}$, 
\beq
{\hat d} = \frac{d u_H^{18/7}}{L_s^{39/7}}e^{\phi_0} \propto \frac{\left(\frac{\mu_0}{L_s}\right)^{9/4}}{L_s^3 T^3} \ ,
\eeq
where high ${\hat d}$ corresponds to high particle density or equivalently to low temperature and vice versa for low ${\hat d}$. Similar interpretations apply to ${\hat k}$ and ${\hat \omega}$. We will begin with the thermodynamic expressions and then move on to the low-energy excitations.

\subsection{Thermodynamics}

We start by direct evaluation of the thermodynamic quantities as calculated in the general framework in Sec.~\ref{setup}. First, the temperature is given by
\beq
T=\frac{\sqrt{\mathcal{B}(u_H)}|\mathcal{F}'(u_H)|}{4\pi}=\sqrt{\frac{11}{12}}\frac{1}{\pi L_s^{1/7}u_H^{6/7}}.
\eeq
The regularized on-shell action is
\bear
-\Omega_{grand}=S^{{\rm on-shell}}_{\rm reg.} &=& -T_7V\Omega_3\int\limits_{0}^{u_H} {\rm d}u\frac{\sqrt{33} e^{-\phi_0}L_s^{54/7}}{7 u^{33/7}}\left(\sqrt{\frac{1}{1+{\hat d}^2}}-1\right)\rc
&=& T_7V\Omega_3\frac{\sqrt{33}{\rm e}^{-\phi_0}L_s^{54/7}}{26 u_H^{26/7}}\left({}_2F_1\left(-\frac{13}{18},\frac{1}{2},\frac{5}{18},-{\hat d}^2\right)-1\right)\rc
&\stackrel{u_H\to \infty}{=}& T_7V\Omega_3\frac{\sqrt{33} { d}^{13/9} e^{4 \phi_0/9} \Gamma \left(\frac{5}{18}\right) \Gamma \left(\frac{11}{9}\right)}{26\sqrt{\pi}L_s^{1/3}} \ ,
\eear
where the last line is the zero-temperature result. The chemical potential is
\beq
\mu = \int\limits_{0}^{u_H} {\rm d}u \frac{d\sqrt{33} }{7\sqrt{d^2+\frac{{\rm e}^{-2\phi_0}L_s^{78/7}}{u^{36/7}}}}\left(\frac{L_s}{u}\right)^{15/7}= \mu_0-L_s\frac{\sqrt{33} \, _2F_1\left(\frac{2}{9},\frac{1}{2};\frac{11}{9};-{\hat d}^{-2}\right)}{8 (u_H/L_s)^{8/7}} \ ,
\eeq
where
\beq
\mu_0 =\frac{\sqrt{11} d^{4/9} {\rm e}^{4\phi_0/9} \Gamma \left(\frac{2}{9}\right) \Gamma \left(\frac{5}{18}\right)}{12 \sqrt{3 \pi }L_s^{1/3}}
\eeq
is the zero-temperature chemical potential. The pressure and energy density expressions at low temperatures are
\bear
&&\!\!\!\!\!\!\!\!\!\epsilon=T_7\Omega_3\left[\frac{\sqrt{33}d^{13/9} e^{4 \phi_0/9}}{52\sqrt{\pi}L_s^{1/3}}  \Gamma \left(\frac{5}{18}\right) \Gamma \left(\frac{2}{9}\right)-\frac{3 \sqrt[6]{\frac{3}{11}} \pi^{4/3}d}{2\ 2^{2/3}}L_s^{7/3}T^{4/3}\right]+\mathcal{O}((L_sT)^{13/3})\,\, \rc
&&\!\!\!\!\!\!\!\!\!p_y=p_x\stackrel{T=0}{=}-\frac{\Omega_{grand}}{V}\,\,\rc
&&\!\!\!\!\!\!\!\!\!p_z\stackrel{T=0}{=}\frac{13}{9}p_x \ ,
\eear
from which we can compute the speed of sound at zero temperature
\beq
c_{s,y}^2=c_{s,x}^2=\frac{4}{9},\quad c_{s,z}^2=\frac{52}{81} \ .
\eeq

%
%
%

\subsection{Low-energy excitations and conductivity}

As for the low energy excitations, we can express the solution to the zero sound equation \eqref{eq:0sound} in terms of the fixed point metric. The 1st order solution and the imaginary correction are
\bear
{\hat \omega^2} &=& \frac{4}{9}{\hat k}_x^2+\frac{2 {\hat d}^{2/9} \Gamma \left(\frac{5}{18}\right) \Gamma \left(\frac{11}{9}\right)}{\Gamma
   \left(\frac{1}{9}\right) \Gamma \left(\frac{7}{18}\right)}{\hat k}_z^2 \ \label{eq:cq}\rc\rc
{\rm Im}\, \delta{\hat \omega}|_{\varphi=0}&=&-\frac{9 \pi^{3/2} \Gamma \left(\frac{20}{9}\right)}{4\sqrt[3]{22} {\hat d}^{4/9} \Gamma \left(\frac{2}{9}\right)^2\Gamma\left(\frac{5}{18}\right) \Gamma \left(\frac{5}{3}\right)^2}{\hat k}_x^{7/3}\, \rc\rc
{\rm Im}\, \delta{\hat \omega}|_{\varphi=\pi/2}&=&-\frac{9 \left(\frac{3}{11}\right)^{2/3} \pi ^{3/2} \Gamma
   \left(\frac{5}{18}\right)^{5/6} \Gamma \left(\frac{20}{9}\right)}{2 \sqrt{2} \sqrt[27]{{\hat d}} \left(\Gamma \left(\frac{1}{9}\right) \Gamma
   \left(\frac{7}{18}\right)\right)^{11/6} \sqrt[6]{\Gamma \left(\frac{11}{9}\right)}
   \Gamma \left(\frac{4}{3}\right)^2}{\hat k}_z^{5/3} \ ,
\eear
where we only report the $x$ and $z$ direction for the imaginary part to avoid cluttering the notation. Notice that the speed of first sound and zero sound in the $x$ direction are equal. On the other hand, the speed of zero sound in the $z$ direction depends on the particle density, making the zero sound fundamentally spatially anisotropic. 

As for the diffusion mode, we have \eqref{eq:diffusion} in terms of the fixed-point metric
\beq
{\hat \omega} = -i\left({\hat k}_x^2 + {\hat k}_z^2\right)\frac{\sqrt{33}\sqrt{1+{\hat d}^2} \, _2F_1\left(\frac{5}{18},\frac{3}{2};\frac{23}{18};-{\hat d}^2\right)}{10} \ . 
\eeq


We can study the behavior of the diffusion in different limits of ${\hat d}$.
\bear
&&D_x\sim \frac{1}{T}, \quad D_z\sim \frac{L_s^{2/3}}{T^{1/3}}, \quad\quad {\hat d} \to 0 \,\,\rc\rc
&&D_x\sim \frac{\mu_0}{(L_sT)^{7/3}}, \quad D_z\sim \frac{\mu_0}{(L_sT)^{5/3}}, \quad\quad {\hat d}\to \infty\ .
\eear

Finally, we express the DC conductivities in terms of this metric. The longitudinal and transverse conductivities in direction $q={\hat x}\cos\chi+{\hat z}\sin\chi$ and $q_{\perp}={\hat x}\sin\chi-{\hat z}\cos\chi$ are
\bear
&\sigma_{qq}=T_7 \Omega _3L_s^{60/7} u_H^{-4/7}e^{-\phi _0}\sqrt{1+{\hat d}^2} \left( \cos ^2\chi+\left(\frac{L_s}{u_H}\right)^{4/7}\sin ^2\chi \right) \,\,,\rc\rc
&\sigma_{qq_{\perp}}=T_7 \Omega _3L_s^{60/7} u_H^{-4/7} \sqrt{1+{\hat d}^2}  \left(1 -\left(\frac{L_s}{u_H}\right)^{4/7}\right)\cos\chi \sin\chi\ .
\eear
Likewise, the behavior of DC conductivities in two different limits are
\bea
&&\sigma_{xx}\sim T_7L_s^{26/3}T^{2/3}, \quad \sigma_{zz}\sim T_7L_s^{28/3}T^{4/3}, \quad\quad {\hat d} \to 0 \,\,\rc
&&\sigma_{xx}\sim \frac{T_7L_s^{41/12}\mu_0^{9/4}}{T^{7/3}}, \quad \sigma_{zz}\sim \frac{T_7L_s^{49/12}\mu_0^{9/4}}{T^{5/3}}, \quad\quad {\hat d} \to \infty\ .
\eea

In the next section, we compare our analytical expressions to numerical ones.

\section{Numerical verification}\label{sec:num}
The numerical methods we use are standard, see, \emph{e.g.}, \cite{Brattan:2012nb}. All of our solutions are obtained at finite temperature.

\subsection{Transition from hydrodynamic to collisionless regime}

In our analytical computations, we saw diffusion and zero sound modes at finite temperatures and zero temperatures, respectively. It is to be expected that, in numerical computations, both modes can appear, and, for different momenta and particle densities, the other is the preferred low-energy mode. The zero sound mode will naturally have to receive some thermal corrections \cite{Bergman:2011rf,Davison:2011ek,Gushterov:2018spg}.

\begin{figure}
	\begin{subfigure}[t]{0.49\textwidth}
	\includegraphics[width=\textwidth]{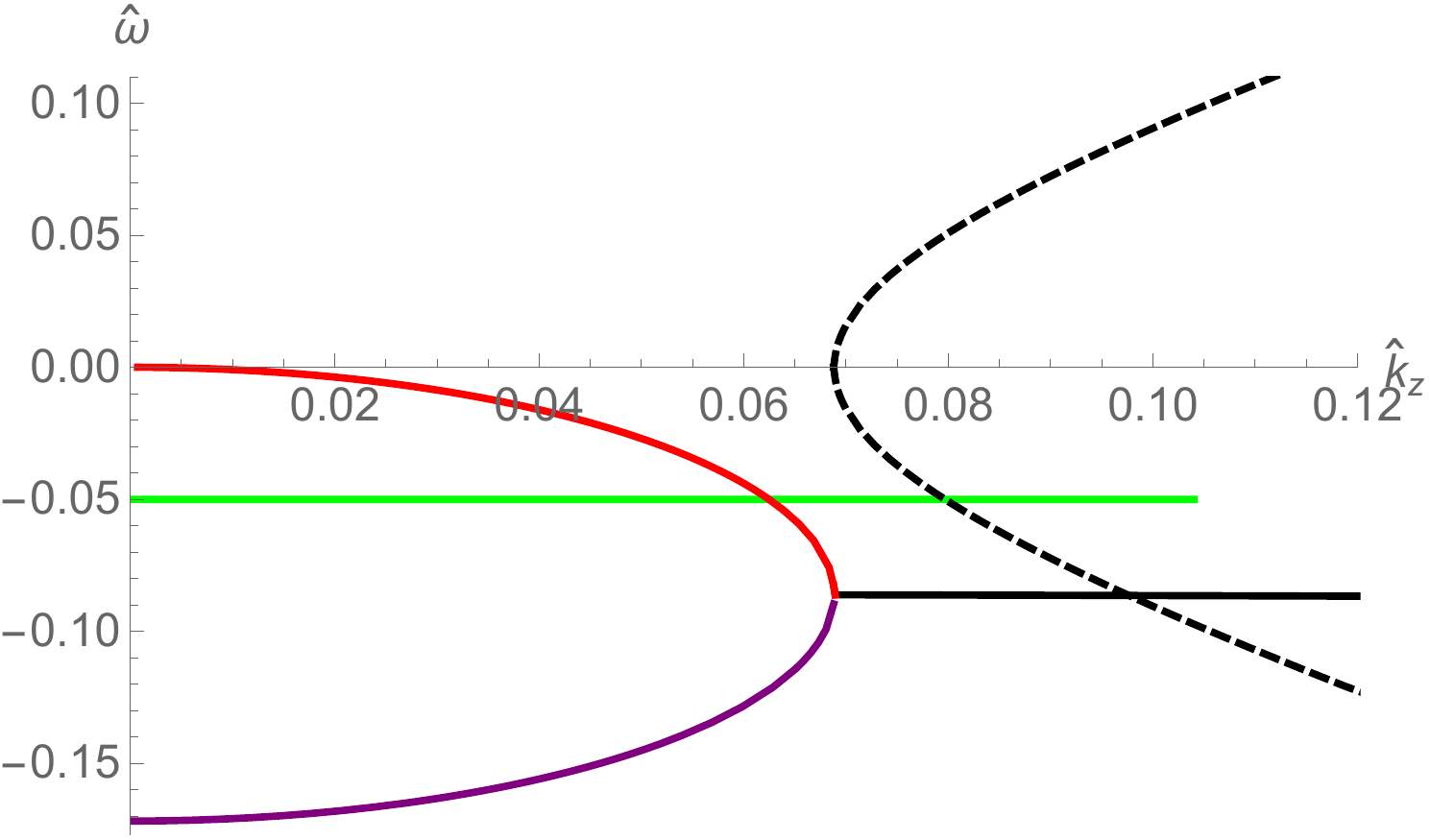}
	\end{subfigure}
	~
        \begin{subfigure}[t]{0.49\textwidth}
	\includegraphics[width=\textwidth]{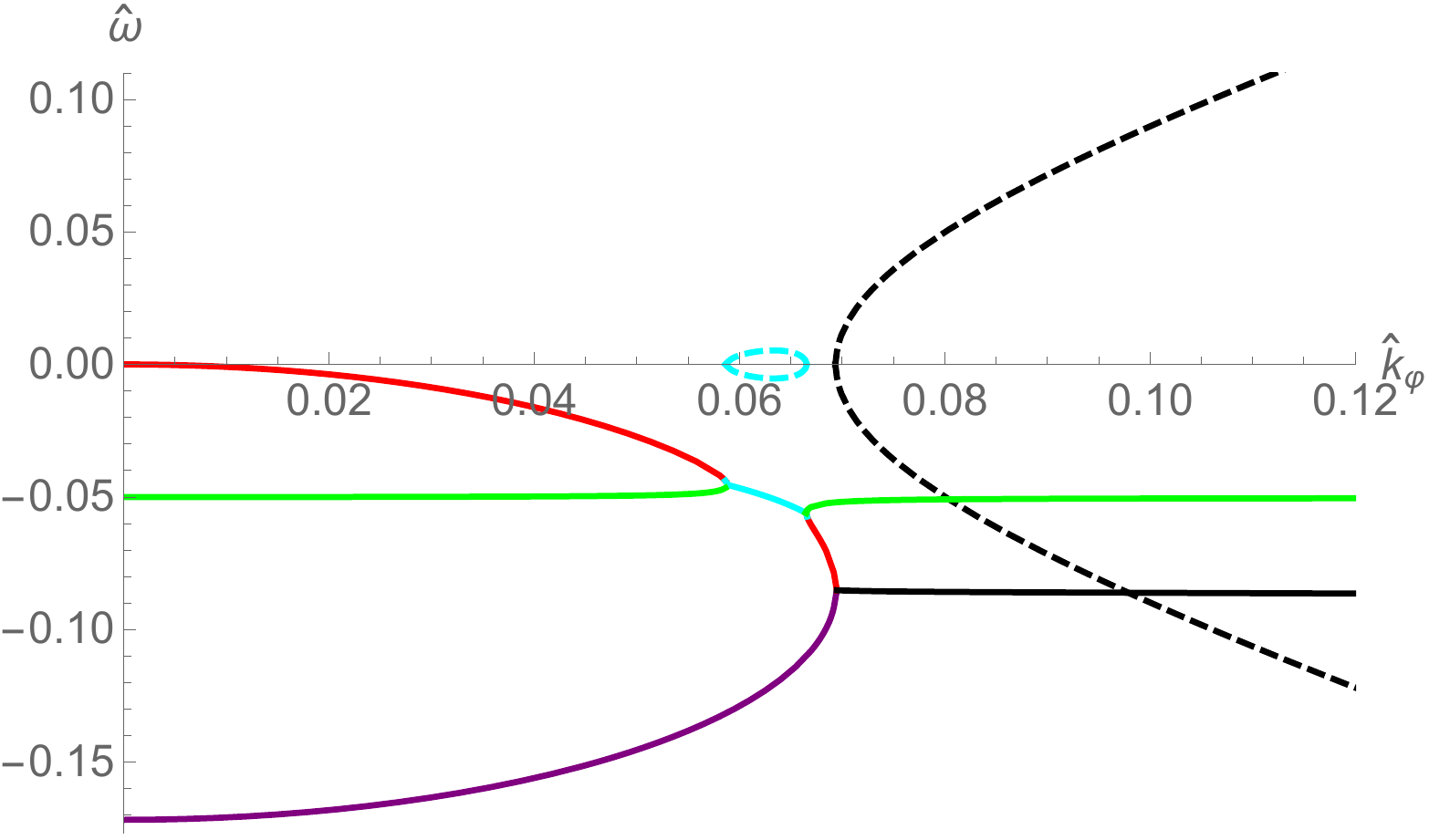}
	\end{subfigure}\\
	\begin{subfigure}[t]{0.49\textwidth}
	\includegraphics[width=\textwidth]{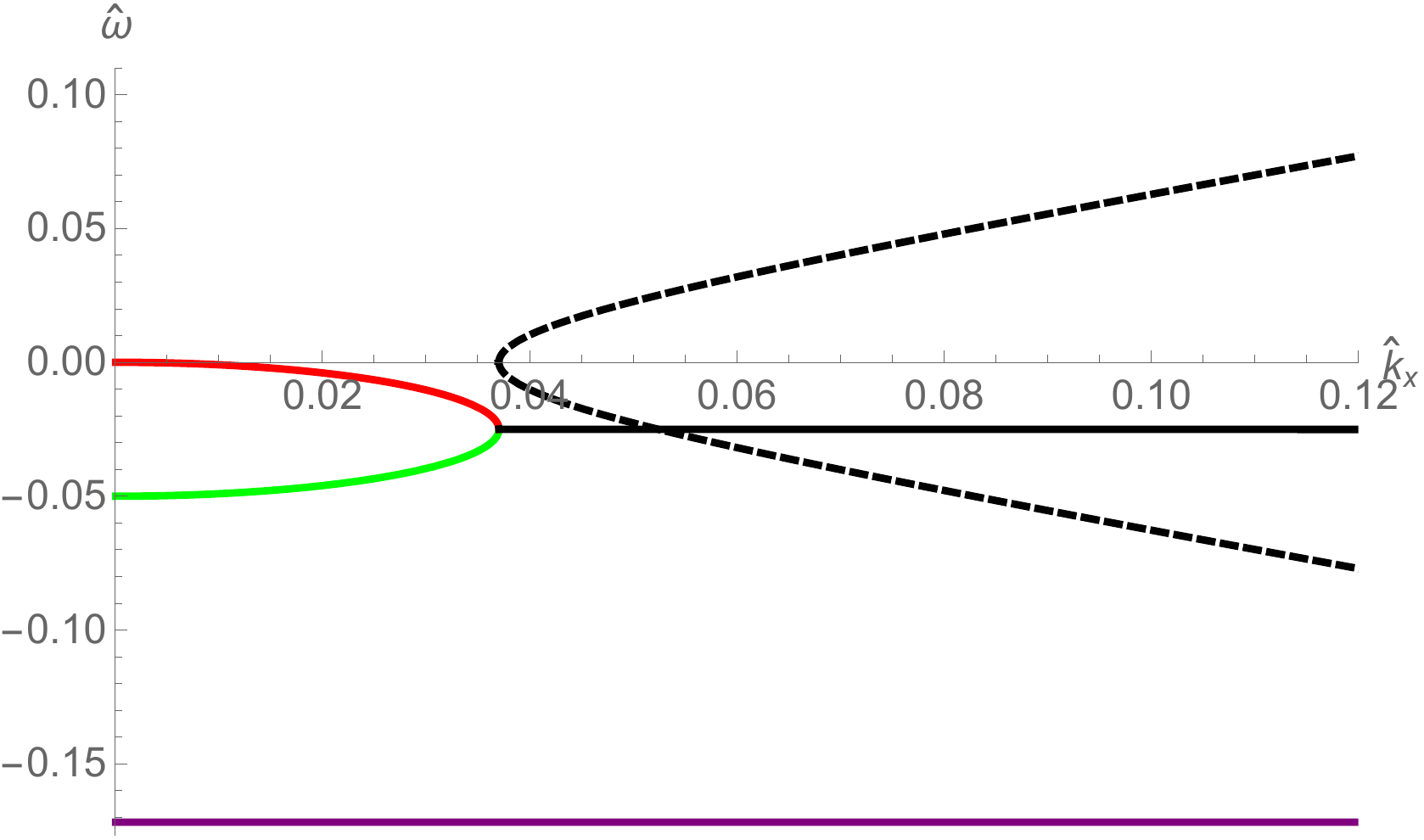}
	\end{subfigure}
	~
        \begin{subfigure}[t]{0.49\textwidth}
	\includegraphics[width=\textwidth]{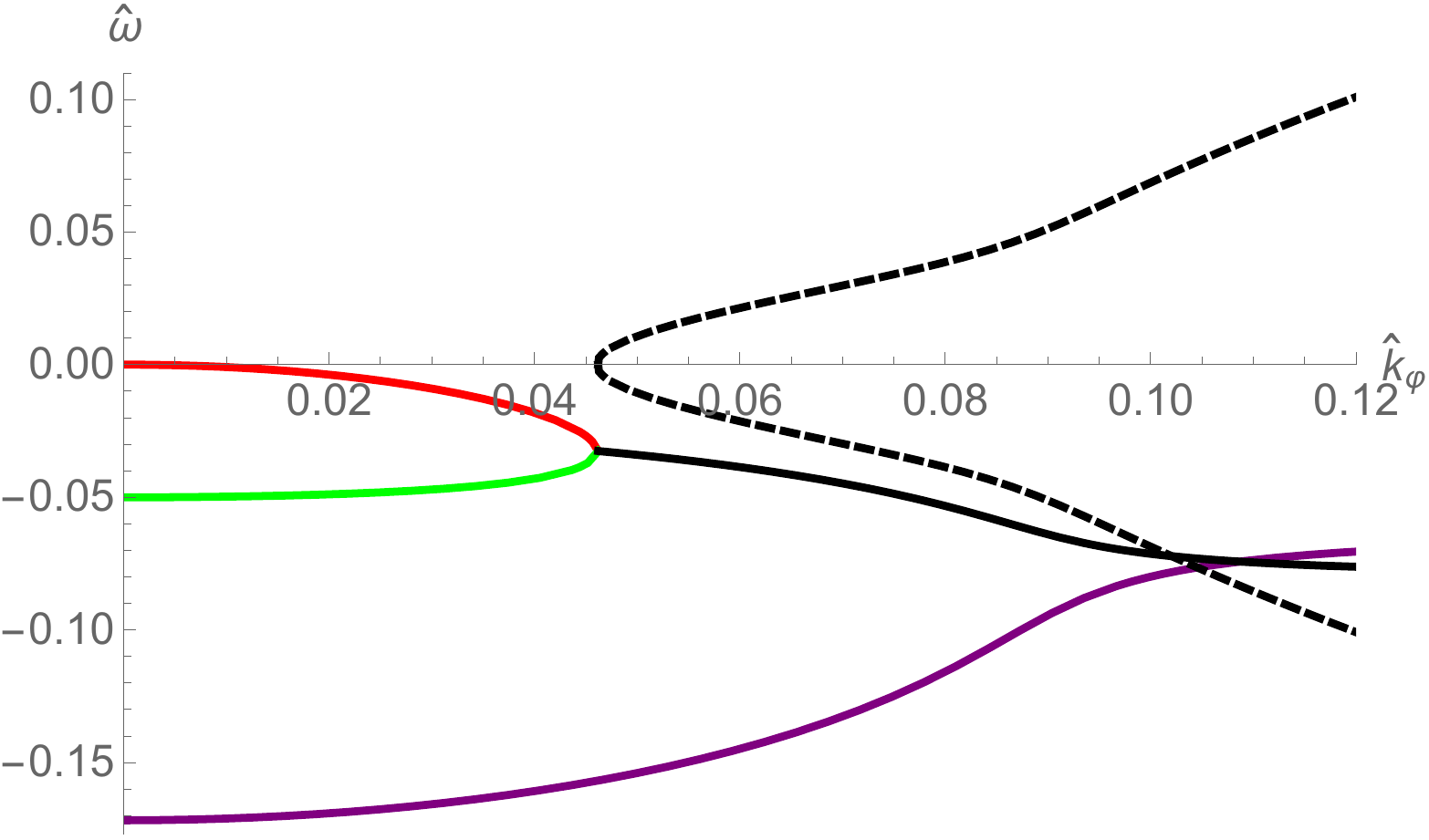}
	\end{subfigure}
\caption{Different modes when ${\hat d}=600$. The dashed lines are real part of ${\hat \omega}$ and the full lines are the imaginary part. Common colors indicate same modes. From top-left corner clockwise, $\varphi=\frac{\pi}{2}$, $\frac{15\pi}{32}$, $\frac{12\pi}{59}$, and, $0$.}
\label{fig:dispplot}
\end{figure}

It turns out that the leading contributions at low momenta come from the diffusion mode. When we scan higher momenta, we see the emergence of a non-zero real part. The higher the particle density or the lower the temperature, the lower the transition momenta. In the numerical analysis, we can also see other purely imaginary modes. See Fig.~\ref{fig:dispplot} for examples. In addition, $x$ and $z$ directions behave differently as the system is fundamentally spatially anisotropic and the speed of zero sound is radically different for the two directions.

\begin{figure}
	\begin{subfigure}[t]{0.49\textwidth}
	\includegraphics[width=\textwidth]{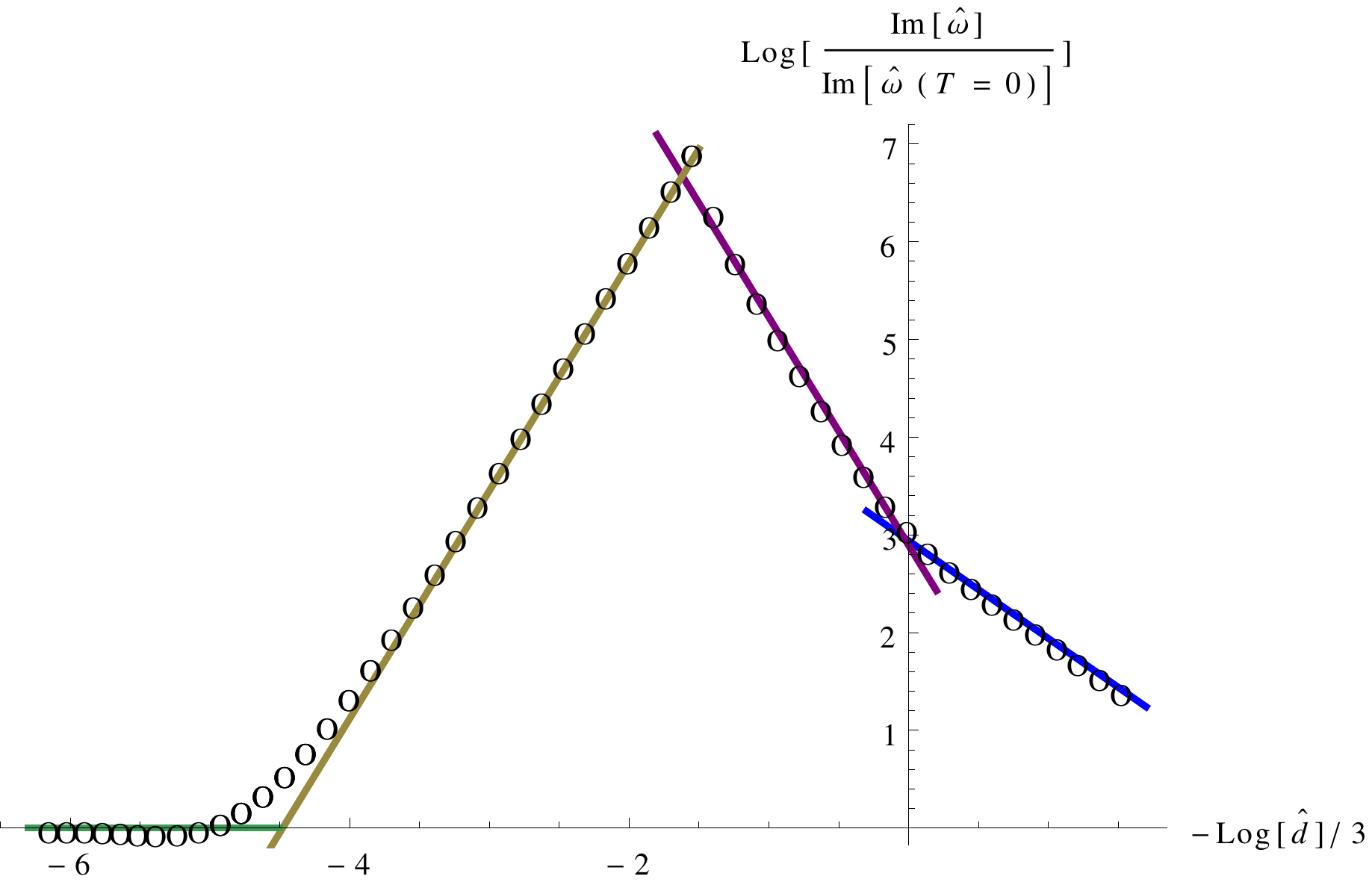}
	\end{subfigure}
	~
	\begin{subfigure}[t]{0.49\textwidth}
	\includegraphics[width=\textwidth]{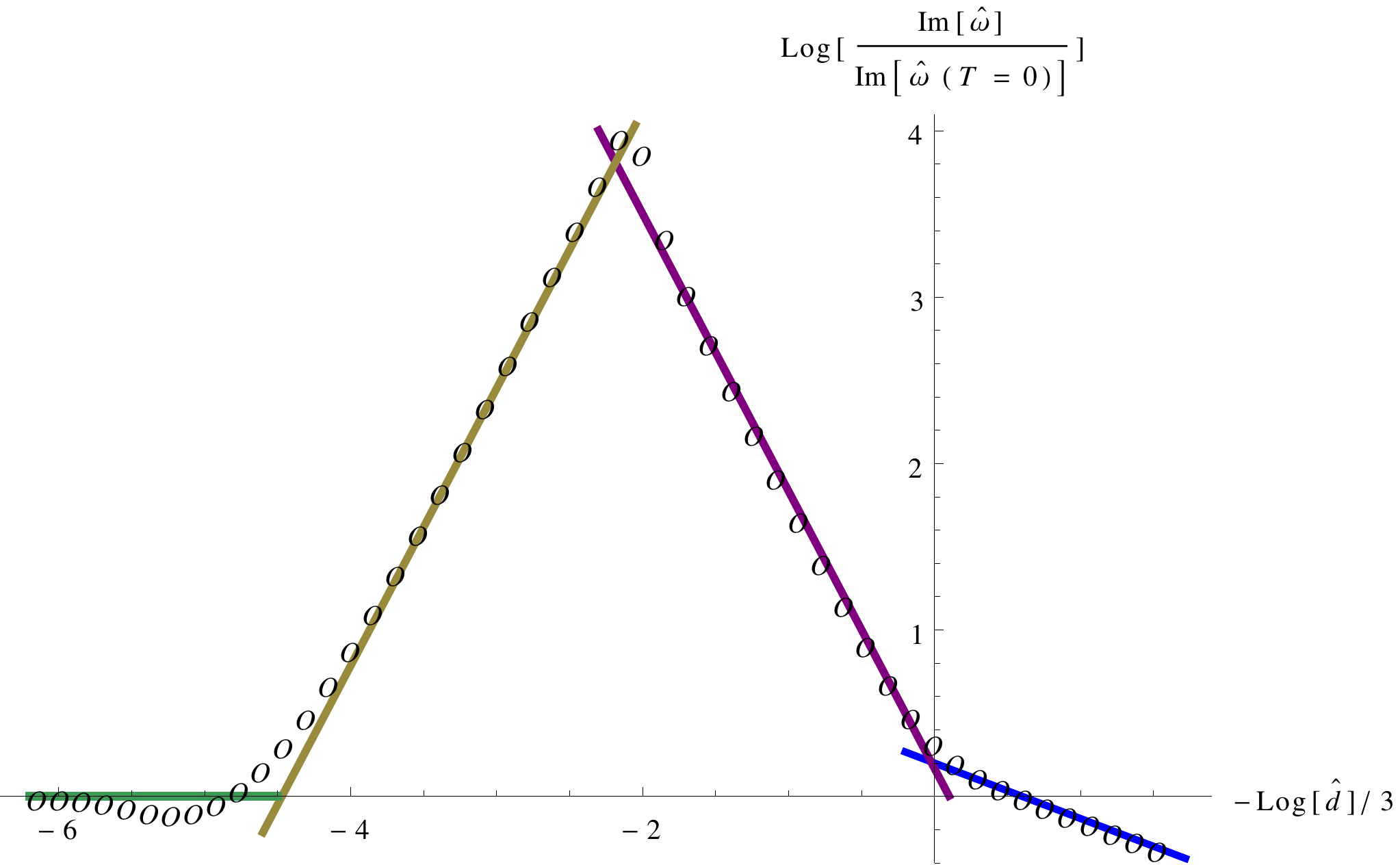}
	\end{subfigure}
\caption{The logarithm of ratio of the analytical zero temperature value and the thermal numeric value of ${\rm Im}\, \omega$. The scaling of $\log {\hat d}$ has been chosen such that $-\frac{1}{3}\log {\hat d}=-\frac{1}{3}\log d+\log T+\mathcal{O}(1)$. The temperature increases to the right. The circles represent the results from our computations and the line segments have been added to help analysis. For the left-hand figure, where $0.02\hat d^{1/3}=\hat k_x\ne 0 = k_z$, the slopes are from the left, $0$, $\frac{7}{3}$, $-\frac{7}{3}$ and $-1$ and for the right-hand figure, where $0.02\hat d^{2/9}=\hat k_z\ne 0 = k_x$, $0$, $\frac{5}{3}$, $-\frac{5}{3}$, and $-\frac{1}{3}$. The slopes conform with our analytic results.}
\label{fig:triangles}
\end{figure}

The real part of ${\hat \omega}$ receives little thermal corrections when it comes to $\nabla_{{\hat k}} {\hat \omega}$. The case is very different for the imaginary part. We can study this more quantitatively by fixing $k$ and $d$ and scaling ${\hat k}$ and ${\hat d}$ appropriately to probe the effect of the temperature. For $k_z=0$, we fix ${\hat k}_x=0.02 {\hat d}^{1/3}$ and, for $k_x=0$, we pick ${\hat k}_z=0.02 {\hat d}^{2/9}$. We compare this to the imaginary part of the zero sound at various ${\hat d}$ to extract the thermal correction to ${\rm Im}\, \omega$. The Fig.~\ref{fig:triangles} contains the logarithmic comparisons of the numerical and analytic values at $T=0$. At low temperatures and high temperatures, the numerical and analytic values agree. At intermediate temperatures, a thermal power law correction can be extracted. For $k_x\neq 0$, the thermal correction is ${\rm Im}\, \omega(T) ={\rm Im}\, \omega(T=0)(1+\gamma_x T^{7/3})$ and for $k_z\neq 0$, it is ${\rm Im}\, \omega(T) ={\rm Im}\, \omega(T=0)(1+\gamma_z T^{5/3})$, where $\gamma$'s are unknown coefficients.

In addition, the point of transition between the hydrodynamic regime and collisionless regime has a simple scaling law for $d\to\infty$. The scaling of critical momentum and angular frequency in the two cardinal directions, $x$ and $z$, are
\bea
&k_{x,c}\sim L_s^{7/3}d^{-4/9}u_H^{-2}\sim\frac{T^{7/3}}{\mu_0}L_s^{7/3},\quad \omega_{x,c}\sim L_s^{7/3}d^{-4/9}u_H^{-2}\sim\frac{T^{7/3}}{\mu_0}L_s^{7/3} \,\,,\rc\rc
&k_{z,c}\sim L_s^{10/7}d^{-1/3}u_H^{-10/7}\sim\frac{T^{5/3}}{\mu_0^{3/4}}L_s^{17/12},\quad \omega_{x,c}\sim L_s^{23/21}d^{-2/9}u_H^{-10/7}\sim\frac{T^{5/3}}{\mu_0^{1/2}}L_s^{7/6} \ .
\eea
We have numerically verified all these behaviors.

\begin{figure}
	\begin{subfigure}[t]{0.32\textwidth}
	\includegraphics[width=\textwidth]{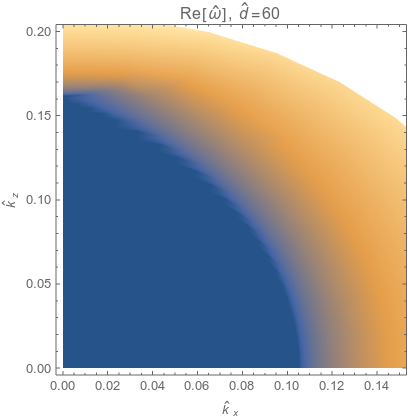}
	\end{subfigure}
	~
	\begin{subfigure}[t]{0.32\textwidth}
	\includegraphics[width=\textwidth]{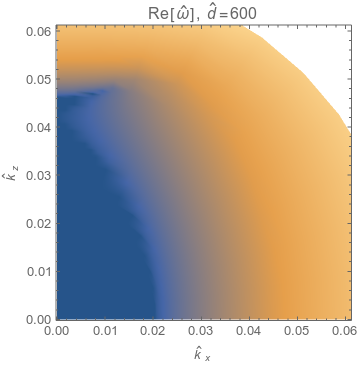}
	\end{subfigure}
	~
	\begin{subfigure}[t]{0.32\textwidth}
	\includegraphics[width=\textwidth]{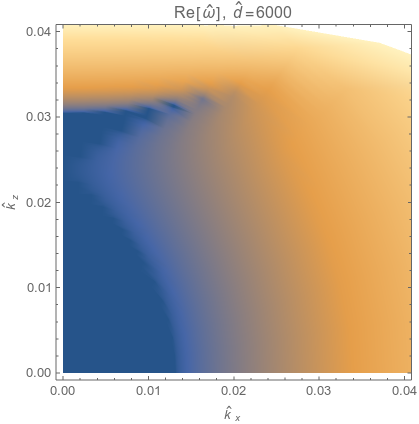}
	\end{subfigure}
\caption{The real part of ${\hat \omega}$ for different ${\hat d}$ scanned at various ${\hat k}_x$ and ${\hat k}_z$. The deep blue indicates the value zero (hydrodynamic regime) while lighter shades are non-zero and positive (collisionless regime). Completely white regions we have not scanned over.}
\label{fig:transitions}
\end{figure}

Additionally, we wish to consider directions other than purely $x$ and $z$. We can scan the low energy modes at various mixed directions and different ${\hat d}$. It turns out that low-temperature and high-density regimes exhibit non-trivial transitioning between hydrodynamic and collisionless regimes. For some small $k_x$, increasing $|k_z|$ might give us a non-zero real part, followed by a vanishing real part, see Fig.~\ref{fig:transitions}. The effect is less prominent for higher temperatures and lower particle densities. The behavior can be better understood by looking at the mixed angle spectra in Fig.~\ref{fig:dispplot}. The transition to zero sound happens at lower momentum. The initial slope is consistent with $c_x\cos\varphi$ from \eqref{eq:cq}. Later on, the zero sound mode transition is complete and the slope is consistent with $c_q$. It is interesting to note that the anisotropy really enter non-trivially in the momentum plane. It would be instructive to perform the same computation in the MT background and check how the non-trivial boundary shape of the hydrodynamic to collisionless regime scales with the anisotropy parameter $a$.

\subsection{Quantitative comparisons}

We scale $u_H$ and $L_s$ out of the conductivity quantities. We can naturally extract the $xx$ and $zz$ components of the (longitudinal) conductivity tensor.
\bear
{\sigma}_{xx}\equiv T_7 \Omega_3L_s^{60/7}{\rm e}^{-\phi_0}\frac{{\hat \sigma}_{xx}}{u_H^{4/7}}\stackrel{\omega=0}{=} T_7 \Omega_3L_s^{60/7}{\rm e}^{-\phi_0}\frac{\sqrt{1+{\hat d}^2}}{u_H^{4/7}} \,\,,
\rc\rc
{\sigma}_{zz}\equiv T_7 \Omega_3L_s^{64/7}{\rm e}^{-\phi_0}\frac{{\hat \sigma}_{zz}}{u_H^{8/7}}\stackrel{\omega=0}{=} T_7 \Omega_3L_s^{64/7}{\rm e}^{-\phi_0}\frac{\sqrt{1+{\hat d}^2}}{u_H^{8/7}}\  .
\eear
The last equalities are for DC conductivity for which we see that, ${\hat \sigma}_{xx}\stackrel{{\hat \omega}=0}{=}{\hat \sigma}_{zz}$.

\begin{figure}[t]
\centering
	\includegraphics[width=0.5\textwidth]{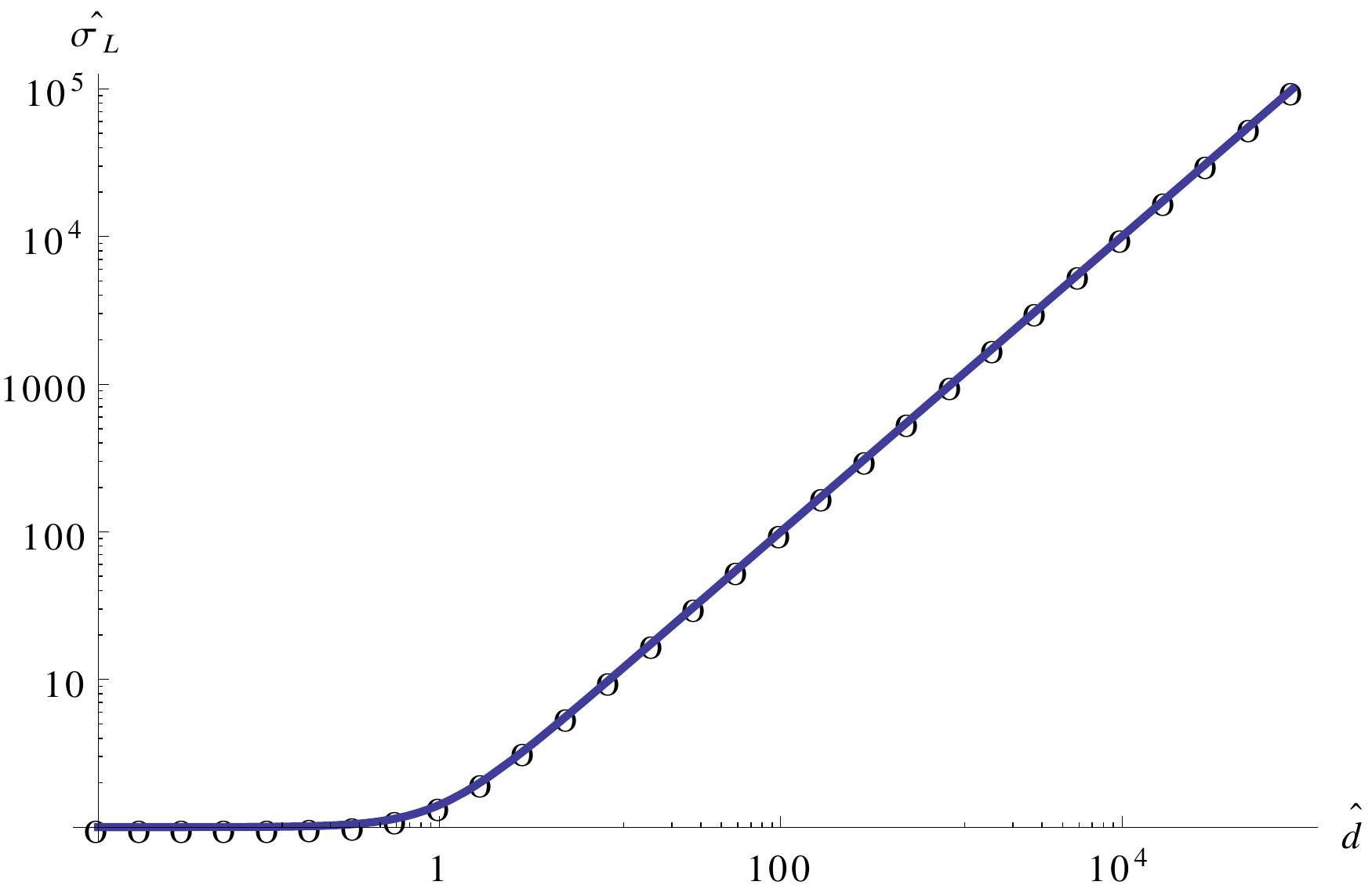}
	\caption{The DC conductivity for various ${\hat d}$. The circles correspond to numerical values while the curve is our prediction for ${\hat \sigma}_{xx}(\omega=0)={\hat \sigma}_{zz}(0)\equiv{\hat \sigma}_L$.}
	\label{fig:dccond}
\end{figure}

\begin{figure}
	\begin{subfigure}[t]{0.45\textwidth}
	\includegraphics[width=\textwidth]{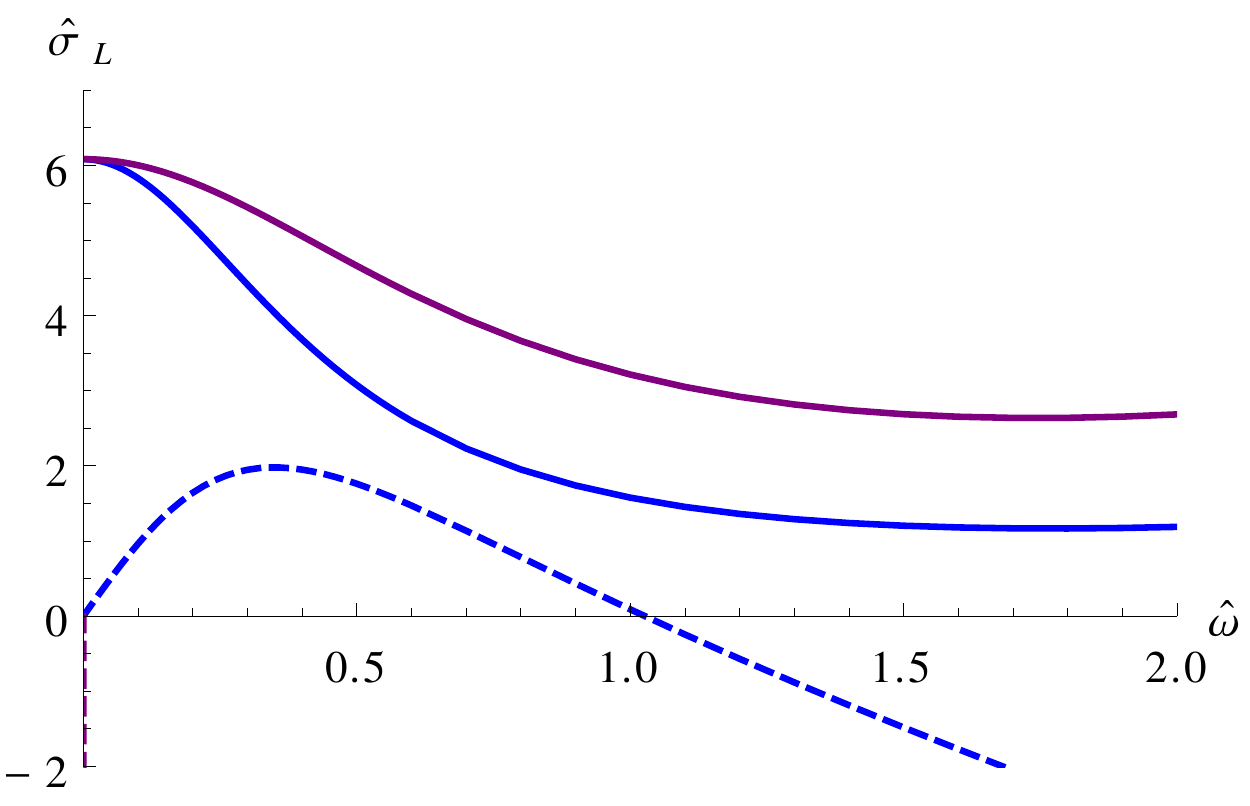}
	\end{subfigure}
	~
	\begin{subfigure}[t]{0.45\textwidth}
	\includegraphics[width=\textwidth]{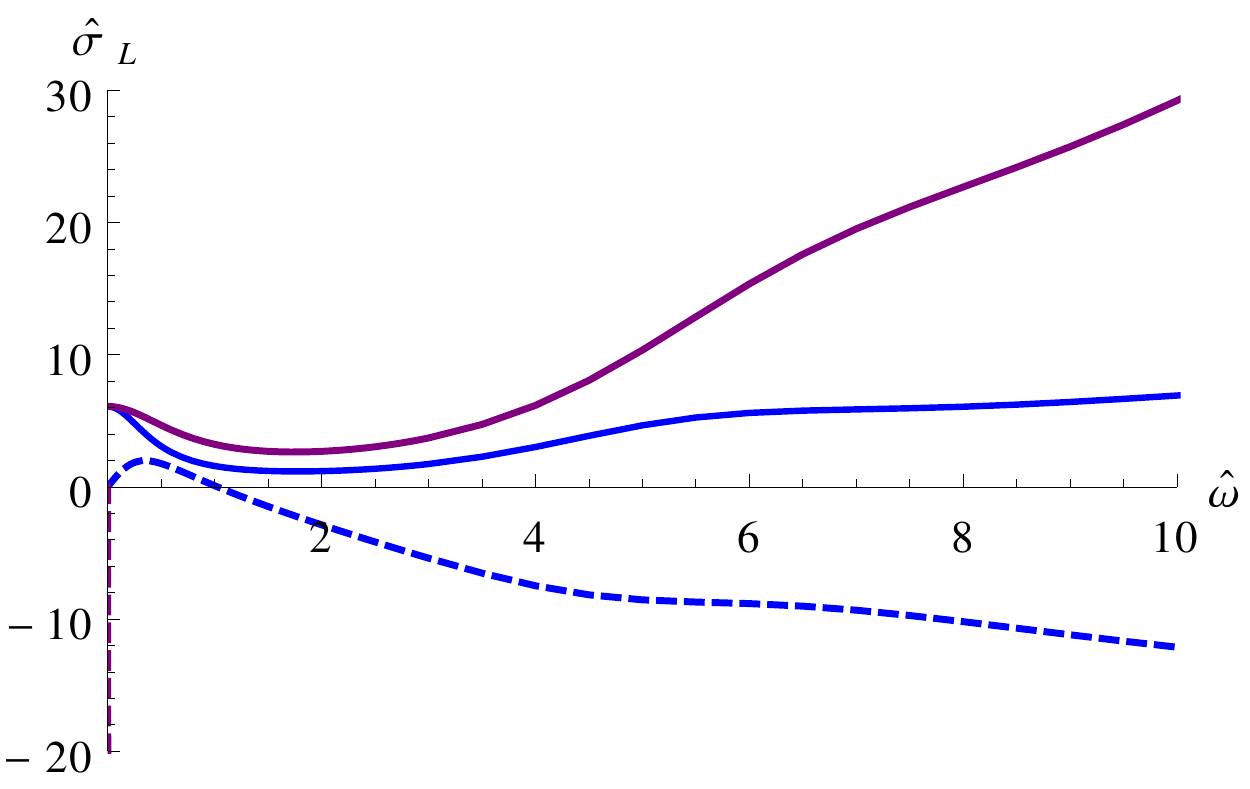}
	\end{subfigure}\\
	\begin{subfigure}[t]{0.45\textwidth}
	\includegraphics[width=\textwidth]{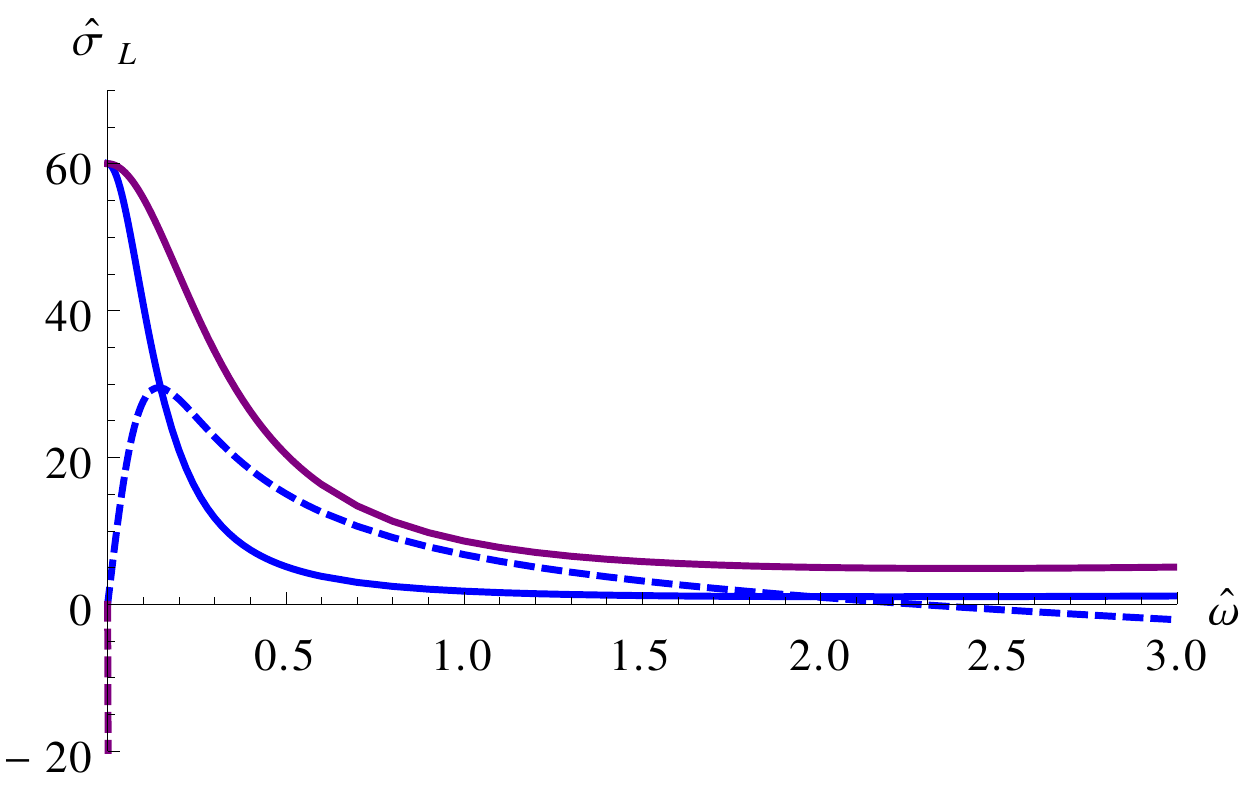}
	\end{subfigure}
	~
	\begin{subfigure}[t]{0.45\textwidth}
	\includegraphics[width=\textwidth]{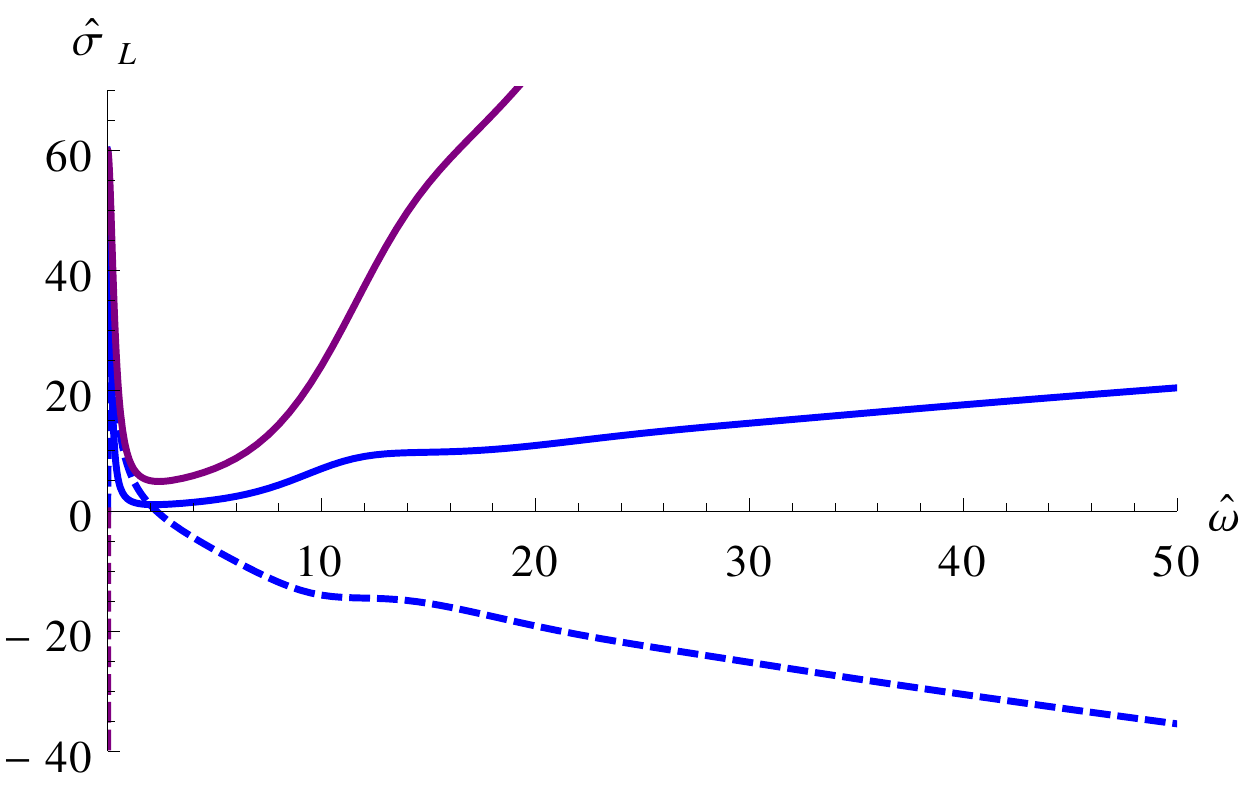}
	\end{subfigure}\\
	\begin{subfigure}[t]{0.45\textwidth}
	\includegraphics[width=\textwidth]{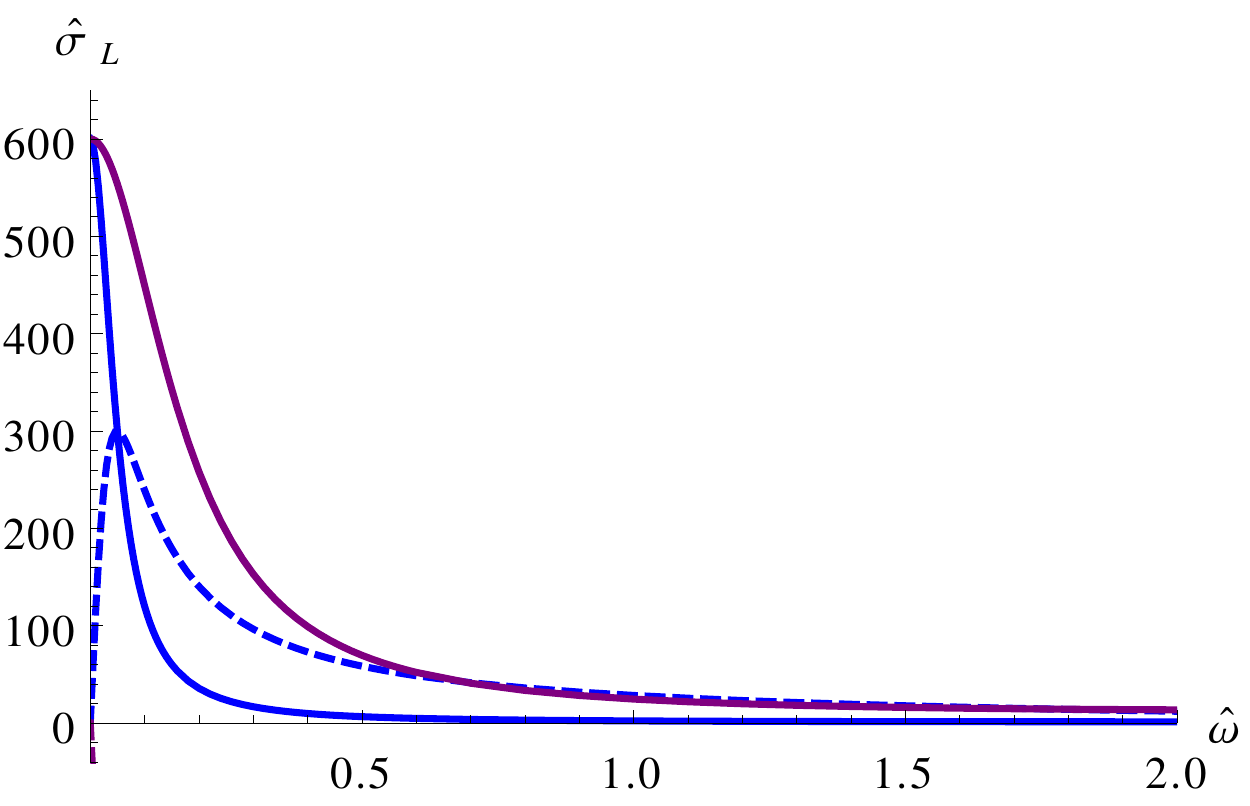}
	\end{subfigure}
	~
	\begin{subfigure}[t]{0.45\textwidth}
	\includegraphics[width=\textwidth]{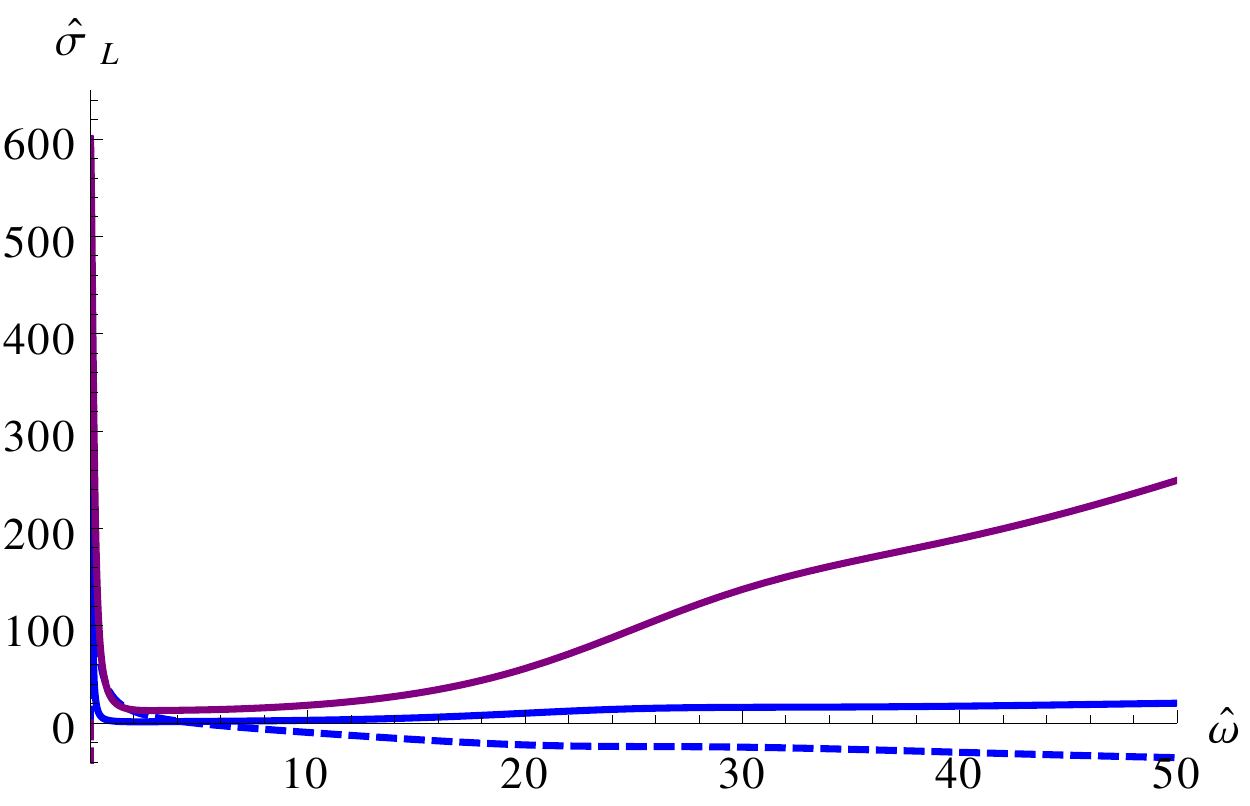}
	\end{subfigure}\\
	\begin{subfigure}[t]{0.45\textwidth}
	\includegraphics[width=\textwidth]{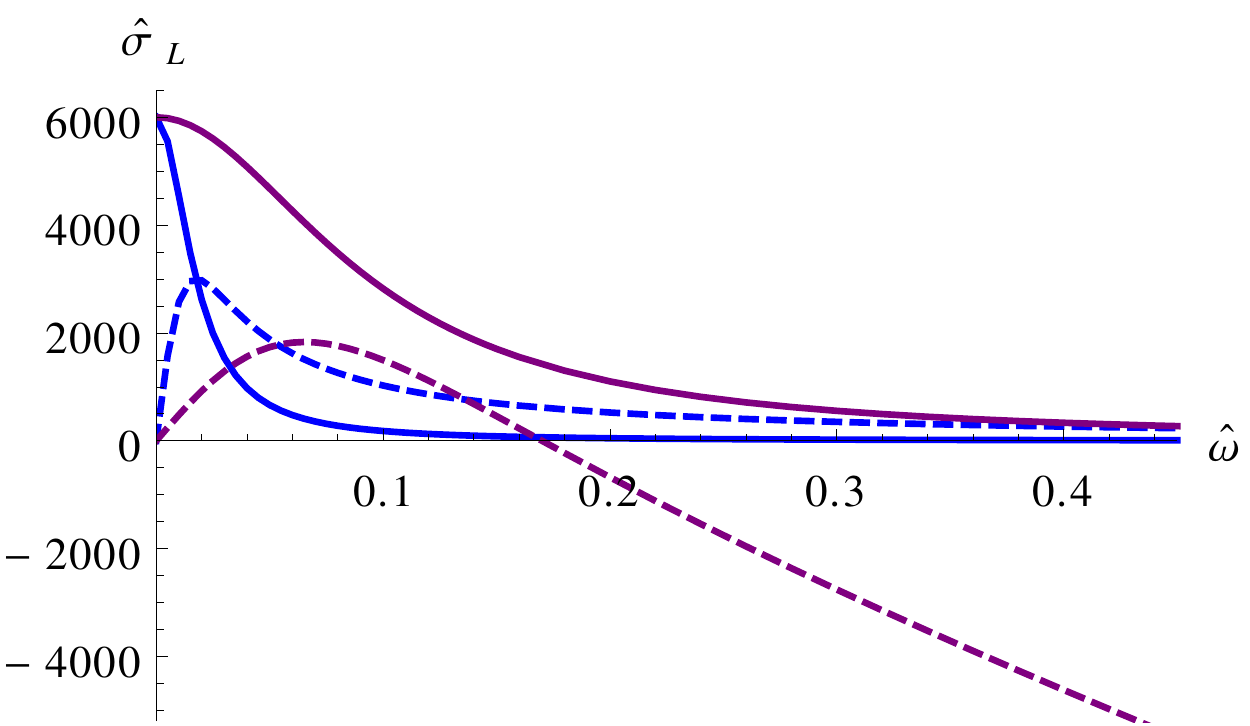}
	\end{subfigure}
	~
	\begin{subfigure}[t]{0.45\textwidth}
	\includegraphics[width=\textwidth]{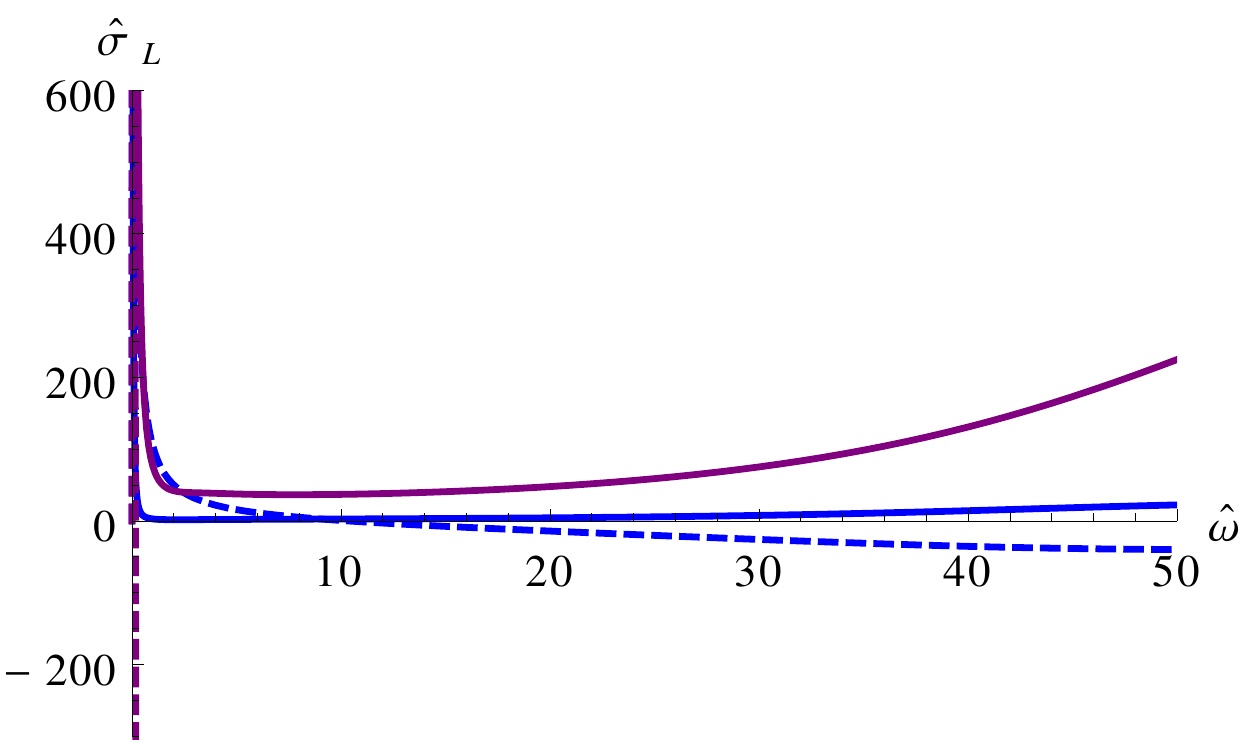}
	\end{subfigure}\\
\caption{The AC conductivity for various ${\hat d}$ as a function of ${\hat \omega}$. Dashed curves represent imaginary parts, solid curves represent real parts of the conductivities. Purple is for the $zz$ component while blue is for the $xx$ component. The figures on the same row have equal ${\hat d}$. From the top row to bottom, ${\hat d}$=6, 60, 600, and 6000. Left panel figure is the zoomed-in version of the right panel, {\emph{i.e.}}, focusing on smaller frequency window.}
\label{fig:accond}
\end{figure}

The numerical analysis confirms the DC conductivity to great accuracy, see Fig.~\ref{fig:dccond}. The low frequency AC conductivity is heavily affected by the thermal effects and our analytical results turn out to be unreliable at high frequencies. The real part of the conductivities start declining at low momenta and then start growing when $\omega\to \infty$ while the imaginary part becomes increasingly more negative. The behavior of ${\hat \sigma}_{zz}$ and ${\hat \sigma}_{xx}$ differ significantly which was to be expected from our analytical computations. The imaginary part of ${\hat \sigma}_{zz}$ becomes negative at very low momenta at low ${\hat d}$ while the imaginary part of ${\hat \sigma}_{xx}$ becomes positive at low momenta but is negative at higher momenta. Fig.~\ref{fig:accond} contains plots of AC conductivity computations.

\begin{figure}
\centering
\includegraphics[width=0.5\textwidth]{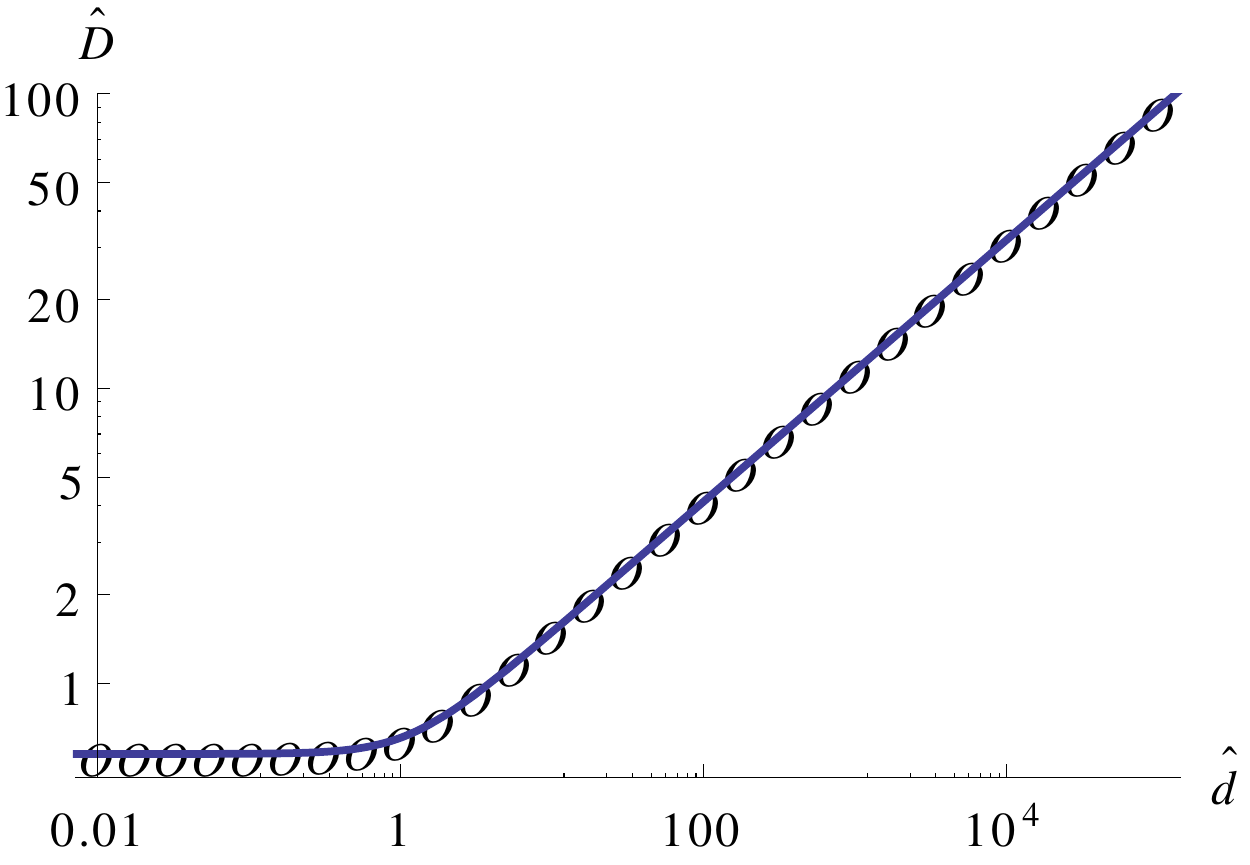}
\caption{The diffusion coefficient for a range of ${\hat d}$. The circles mark the numerical values while the solid curve is our analytical prediction.}
\label{fig:diffplot}
\end{figure}

When we scaled out $u_H$, the diffusion coefficient for low momenta became equal for all directions. We find excellent agreement between the numerical and analytical value, as depicted in Fig.~\ref{fig:diffplot}.

\section{Conclusions and outlook}\label{sec:conc}

In this paper we performed an in-depth study of flavor physics in an anisotropic background. We investigated both the thermodynamics of massless fundamental degrees of freedom as well as their dynamical properties in a cold and dense environment. One of our intermediate goals was to fill in an important gap in the literature as anisotropic backgrounds have become more and more under the scope of recent discussions. The results that we delivered can be directly used in many contexts since we kept the metric components general. 

A long-term goal and definitely a much more ambitious one is to identify the physical system that could be genuinely modeled by the anisotropic setups we studied here. While the original Mateos-Trancanelli background was fairly well suited in studying heavy ion physics, here the emphasis has been on dense and cold regimes, naturally reached inside neutron stars. For example, perturbations of the self-gravitating fluids will induce anisotropic stresses. The local anisotropy of energy density and pressure may then even lead to cracking of the star \cite{Herrera:1992lwz,Herrera:1997plx,Herrera:2015vca}. If anisotropy is present in the ultradense regime in neutron stars, it will have significant effects on the stellar properties and structures. 

Recently, the first steps have been taken in holography to model dense and cold quark matter phase that could be realized inside neutron stars \cite{Hoyos:2016zke,Annala:2017tqz}. A generalization of such a study by incorporating the anisotropy in the strongly coupled quark matter phase would be interesting. This might potentially pave the way forward in understanding if anisotropies are relevant in compact objects. We hope to return to this issue in the future.

\paragraph{Acknowledgments}

\noindent
We would like to thank Matteo Baggioli for an interesting comment.
The work of G. ~I. is supported by FAPESP grant 2016/08972-0 and 2014/18634-9.
N.~J. is supported in part by the Academy of Finland grant no. 1303622.
J.~J. is in part supported by the Academy of Finland grant no. 1297472 and the U. Helsinki Graduate School PAPU.
 A. ~V. ~R.   is funded by the Spanish grants FPA2014-52218-P and FPA2017-84436-P by Xunta de Galicia (GRC2013-024),  by FEDER and by  the Maria de Maeztu Unit of Excellence MDM-2016-0692.

\appendix

\vskip 1cm
\renewcommand{\theequation}{\rm{A}.\arabic{equation}}
\setcounter{equation}{0}

\section{Some useful integrals}\label{appendix:calculations}
Let us collect in this appendix some integrals which are useful in the analysis of the collective excitations of the matter in case of the fixed point Lifshitz-like metric. First of all, we define the integral $I_{\lambda_1,\lambda_2}(u)$ as:
\beq
I_{\lambda_1,\lambda_2}(u)\,\equiv\,\int_{0}^{u}\,\frac{
{\tilde u}^{\lambda_1}\,d{\tilde u}}{ ({\tilde u}^{\lambda_2}+d^2)^{\frac{1}{2}}}\,\ .
\label{I_lambda12_def}
\eeq
This integral can be explicitly performed in terms of the hypergeometric function:
\beq
I_{\lambda_1,\lambda_2}(r)\,=\,\frac{2}{2+2\lambda_1-\lambda_2}\,
u^{1+\lambda_1-\frac{\lambda_2}{ 2}}\,
F\Big(\frac{1}{ 2}, \frac{1}{ 2}-\frac{\lambda_1+1}{\lambda_2};\frac{3}{ 2}-\frac{\lambda_1+1}{\lambda_2}
;-\frac{d^2}{ u^{\lambda_2}}\Big)\, .
\label{I_lambda12_value}
\eeq
For large $u$, assuming that $\lambda_2$ and $\lambda_1+1$ are negative, we have the expansion:
\beq
I_{\lambda_1,\lambda_2}(u)=-\frac{1}{\lambda_2} B\left(\frac{\lambda _1+1}{\lambda _2},\frac{1}{2}-\frac{\lambda _1+1}{\lambda _2}\right)d^{2\frac{\lambda_1+1}{\lambda _2}-1}\,+\frac{u^{\lambda _1+1}}{d(\lambda _1+1)}+\ldots \ .
\label{I_lambda12_expansion}
\eeq
Let us next define $J_{\lambda_1,\lambda_2}(u)$ in the form:
\beq
J_{\lambda_1,\lambda_2}(u)\,\equiv\,\int_{0}^{u}\,
\frac{{\tilde u}^{\lambda_1}\,d{\tilde u}}{({\tilde u}^{\lambda_2}+d^2)^{\frac{3}{ 2}}}\, ,
\label{J_integral_definition}
\eeq
which can also be computed explicitly:
\beq
J_{\lambda_1,\lambda_2}(u)\,=\,\frac{2}{2+2\lambda_1- 3\lambda_2}\,
u^{1+\lambda_1-\frac{3\lambda_2}{ 2}}\,
F\Big(\frac{3}{ 2}, \frac{3}{ 2}-\frac{\lambda_1+1}{\lambda_2};\frac{5}{ 2}-\frac{\lambda_1+1}{\lambda_2}
;-\frac{d^2}{ u^{\lambda_2}}\Big)\, .
\label{J_value}
\eeq
For large $u$, when  $\lambda_2$ and $\lambda_1+1$ are both negative, we can expand 
$J_{\lambda_1,\lambda_2}(u)$ as:
\beq
J_{\lambda_1,\lambda_2}(u)=-\frac{1}{ \lambda_2}\,
B\Big(\frac{\lambda_1+1}{ \lambda_2}, \frac{3}{ 2}-\frac{\lambda_1+1}{ \lambda_2}\,
\Big)\,d^{2\frac{\lambda_1+1}{ \lambda_2}-3}\,+\,\frac{u^{\lambda_1+1}}{ (\lambda_1+1)d^3}+\ldots \ .
\eeq

\end{document}